\documentclass{aa}  
\usepackage{adjustbox}
\usepackage{float}
\usepackage{xcolor}
\usepackage[dvipsnames]{xcolor}

\usepackage{graphicx}
\usepackage{txfonts}
\usepackage{lipsum}
\usepackage{subcaption}
\usepackage{lscape}
\usepackage{placeins}

\usepackage{url}     
\usepackage{xurl}

\usepackage{txfonts}
\usepackage{enumitem}
\setlist[itemize]{label=\textbullet}
\usepackage[]{hyperref}
\usepackage{cleveref}[2012/02/15]
\crefformat{footnote}{#2\footnotemark[#1]#3}
\usepackage{xcolor}
\usepackage[dvipsnames]{xcolor}
\hypersetup{
    colorlinks=true,
    linkcolor=blue,
    citecolor=blue,
    filecolor=blue,      
    urlcolor=blue,
    }
\usepackage{morefloats}
\begin{document}
    \title{AVICA: A fully automated CASA pipeline for large volume VLBI data calibration
        \thanks{\texttt{AVICA} is available as a Python package at \url{https://pypi.org/project/avica}}
        }
   \subtitle{}

\author{A. Kumar\inst{1,2}\corrauth{avialxee@gmail.com}
          \and C. Casadio\inst{1,2}
          \and M. Janssen\inst{3}
          \and D. \'Alvarez-Ortega\inst{1,2}
          \and F. M. P{\"o}tzl\inst{1,2}
          }
   \institute{Institute of Astrophysics, Foundation for Research and Technology – Hellas, N. Plastira 100, Voutes GR-70013, Heraklion, Greece
       \and Department of Physics, University of Crete, 70013 Heraklion, Greece
       \and Department of Astrophysics, Institute for Mathematics, Astrophysics and Particle Physics (IMAPP), Radboud University, P.O. Box 9010, 6500 GL Nijmegen, The Netherlands
        }
  \abstract
   {Calibrating large volumes of Very Long Baseline Interferometry (VLBI) data is a time-consuming process, traditionally requiring significant human intervention at every stage, from data inspection and parameter tuning to calibrator selection. While the Common Astronomy Software Applications (\texttt{CASA}) package is the standard data reduction tool across major radio observatories, no existing \texttt{CASA}-based pipeline is capable of operating in a fully automated manner across the heterogeneous data formats produced by the Very Long Baseline Array (VLBA) over three decades of operations. The Search for Milli-Lenses (SMILE) project, requiring the calibration of $\sim5000$ VLBA sources, makes such blind automation a practical necessity.}
   {We introduce the Automated VLBI pipeline in \texttt{CASA} (\texttt{AVICA}), designed to automate the calibration of VLBA datasets for the SMILE sample while making its functionalities available for the broader VLBI community.}
   {\texttt{AVICA} extends the existing \texttt{CASA}-based \texttt{rPICARD} calibration framework by automating the preprocessing of archival VLBA data, the selection of calibrators and reference antennas, and the execution of the full calibration workflow. Preprocessing operates on both the FITS-IDI and Measurement Set (MS) data formats, extracting only the required sources from large archive files to reduce data volume and loading time. Calibrators and reference antennas are ranked automatically using signal-to-noise ratio (S/N) from the FFT-based fringe detection, and the resulting parameters are passed to \texttt{rPICARD} for calibration. Workflow management and progress tracking are handled by Automated Logical Framework for executing Dynamic scripts (\texttt{ALFRD}), which orchestrates pipeline execution for each dataset and records results to an external spreadsheet in real time.}
   {\texttt{AVICA} was validated on a sample of 1000 sources with NRAO archival data spanning observations from 1995 to 2023, for a total of $1372$ individual band-separated observations across the S, C, X, U, and K bands. The pipeline successfully produced calibrated output for 978 sources ($97.8\%$), with the $22$ failures attributable to corrupted or incomplete input data. The mean per-source execution time across the 1000-source sample was approximately 30 minutes, using MPI parallelization with up to 20 cores.}
   {\texttt{AVICA} demonstrates that fully blind calibration of heterogeneous archival VLBA data is achievable using \texttt{CASA}, without manual parameter input. Although validated on archival VLBA data, the underlying algorithms are designed to be generalizable to other VLBI arrays, and the automated calibrator and reference antenna selection will be incorporated into a future \texttt{rPICARD} release, extending blind-automated calibration to any supported array. \texttt{AVICA} and \texttt{ALFRD} are available as open-source Python packages.}

   \maketitle
\nolinenumbers

\section{Introduction}
Very Long Baseline Interferometry (VLBI) is a technique that combines signals from radio antennas separated by intercontinental baselines, achieving up to micro-arcsecond angular resolution. The calibration of VLBI data is a multi-step process requiring corrections for atmospheric delays, instrumental effects, and clock offsets between independently recording stations. For a single observation, this process is time intensive and traditionally requires days of expert effort, involving careful inspection of the data, manual selection of calibrators and reference antennas, and iterative tuning of calibration parameters.
The Very Long Baseline Array (VLBA), operated by the National Radio Astronomy Observatory (NRAO), is a dedicated ten-station VLBI array that has been in continuous operation since 1994, accumulating a large public archive of heterogeneous observations spanning multiple frequencies. Historically, VLBA data reduction has been performed with the Astronomical Image Processing System (\texttt{AIPS}) \citep[][]{Greisen2003}, which provides robust handling of the native VLBA data formats. However, the Common Astronomy Software Applications (\texttt{CASA}) package \citep[]{McMullin2007, CASA2022} has emerged as the standard reduction environment across major radio observatories, including the Karl G. Jansky Very Large Array (VLA) and Atacama Large Millimeter/submillimeter Array (ALMA), offering an actively developed framework with modern data formats, an interactive imaging interface, and a growing suite of VLBI-capable tasks \citep{bemmelCASAonfringe2022}. Notably, \texttt{CASA}-based Radboud PIpeline for the Calibration of high Angular Resolution Data (\texttt{rPICARD}) has been adopted as a standard calibration pipeline for the Event Horizon Telescope (EHT) \citep{Janssen2019}.

Migrating VLBA calibration workflows to \texttt{CASA}, therefore, enables access to this broader ecosystem, with the \texttt{rPICARD} demonstrated to outperform conventional manual calibration methods \citep{Kim2023}. However, processing the full heterogeneity of the archival VLBA data spanning three decades of evolving correlator formats, recording bandwidths, and file structures remains a significant challenge at scale.

Pipelines for automating VLBI data calibration to date include \texttt{rPICARD} for \texttt{CASA} and \texttt{VIPCALs} \citep{VIPCALs2025}, EHT-HOPS \citep{Blackburn2019}, the Korean VLBI Network (KVN) pipeline \citep{Hodgson2016}, the European VLBI Network (EVN) pipeline \citep{Reynolds2002}, and VLBARUN for the \texttt{AIPS} community. While these pipelines have reduced the manual effort required for individual datasets, they remain driven by configuration files that require the user to supply observation-specific input parameters. An exception is \texttt{VIPCALs}, which automates calibrator and reference antenna selection for VLBA data within \texttt{AIPS}. While an experienced radio astronomer can fine-tune these parameters for a single dataset, the approach becomes impractical at the scale of large projects involving hundreds to thousands of heterogeneous observations.

The Search for Milli-Lenses (SMILE) project \citep{casadio2021} aims to constrain the number density of Supermassive Compact Objects (SMCO) with masses in the range $10^6$--$10^9\,M_\odot$, improving on previous limits \citep{Wilkinson2001} by more than an order of magnitude. The detection of milli-lens candidates produced by SMCOs such as dark matter subhaloes relies on the identification of characteristic double or multiple compact components in high-resolution VLBI images \citep{Potzl2025}. The SMILE sample comprises $\sim5000$ radio-loud sources, constituting a flux-limited sub-sample (flux density $>50$\,mJy at 8.4\,GHz) from the Cosmic Lens All Sky Survey \citep[CLASS;][]{Myers2003} carried out with the VLA. VLBA data for all SMILE sources are either available in the NRAO archive\footnote{\url{https://data.nrao.edu/}} or were recently obtained as part of the project. The scale of the sample makes manual calibration infeasible and provides the primary motivation for the development of a fully automated pipeline.

In this work, we present the Automated VLBI pipeline in \texttt{CASA} (\texttt{AVICA}), a modular, fully automated calibration pipeline for batch processing VLBA data using \texttt{CASA}. Once configured, \texttt{AVICA} requires only the target source name and the path to the input data, automating all intermediate steps from raw data preprocessing to the output of a calibrated visibility file. The pipeline builds on \texttt{rPICARD} for the core calibration scheme, extending it with automated preprocessing calibrator and reference antenna selection. Workflow management and progress tracking are handled by Automated Logical Framework for executing Dynamic scripts (\texttt{ALFRD}), a scheduling backend that coordinates execution across large dataset batches and records results in real time to an external spreadsheet.

The paper is structured as follows. Section~\ref{sec:codestructure} describes the code structure of the \texttt{AVICA} pipeline. Section~\ref{sec:calibrationvlbi} outlines the VLBI calibration framework and fringe fitting formalism. Section~\ref{sec:workflow} describes the pipeline workflow in detail. Section~\ref{sec:results} presents the results and discussion of the validation on archival VLBA datasets for 1000 individual sources, including pipeline performance, a comparison against manually calibrated data, and future developments. Conclusions are presented in Section~\ref{sec:summary}.

\section{Code structure of \texttt{AVICA}}\label{sec:codestructure}

\texttt{AVICA} is designed with two levels of usability. By default, \texttt{AVICA} operates as a fully automated end-to-end pipeline with \texttt{ALFRD} bundled as the workflow backend. The individual components of the \texttt{AVICA} Python package can also be invoked independently, giving experienced users full control over pipeline functionalities such as calibration methodology, preprocessing steps, and workflow logic. For executing calibration steps, \texttt{AVICA} uses \texttt{rPICARD} as the core calibration framework.

\subsection{\texttt{AVICA} Python package}\label{sec:codeavica}

\texttt{AVICA} Python package provides an extensive library for manipulating and inspecting both FITS-IDI and Measurement Set (MS) file formats, making use of the \texttt{CFITSIO} \citep{CFITSIO1999} and \texttt{Astropy} \citep{Astropy2013} libraries. Before calibration is invoked, \texttt{AVICA} inspects the raw visibility data, preprocesses the data and assembles the ancillary metadata required for the subsequent workflow. These preprocessing steps operate directly on the uncalibrated data and are a prerequisite for reliable automated calibration across the heterogeneous NRAO archive. The steps are described in detail in Sects.~\ref{sec:preprocessFITSIDI} and~\ref{sec:preprocessMS}.

\texttt{AVICA} is configured through a text file or via command line arguments, specifying parameters such as the output directory and the root folder containing the raw visibility files. A minimal setup requires only the source name and the path to the raw visibility file. All other parameters are initialized to array-specific default values. Every parameter is individually overridable, allowing the pipeline to be fully adapted to any array and observation setup. Full details of the available configuration options and installation instructions are provided in the documentation\footnote{\label{avicadocumentation}\url{https://avica.readthedocs.io/en/latest/}}.

\subsection{\texttt{ALFRD}: Workflow manager}\label{sec:codealfrd}
\texttt{ALFRD}\footnote{\url{https://github.com/avikhagol/alfrd}} is the scheduling and workflow management module bundled with \texttt{AVICA}.
While integrated into the pipeline, \texttt{ALFRD} is developed as a standalone Python package that can be installed and utilized independently for broader task-scheduling applications. 
Commonly used workflow managers such as \texttt{Snakemake}, \texttt{Nextflow} or \texttt{Airflow} offer similar orchestration capabilities, but each introduces a substantial dependency stack of its own. This is problematic in a \texttt{CASA} environment, where the calibration software pins specific versions of shared libraries such as \texttt{protobuf}, and where introducing a heavy external manager risks version conflicts that can break the runtime. \texttt{ALFRD} was therefore developed in-house to provide the orchestration required by the pipeline as a lightweight Python package with minimal dependencies, avoiding any interference with the \texttt{CASA} software stack while keeping installation and configuration simple for the end user. It is designed to sustain uninterrupted execution across large dataset batches, featuring the ability to resume workflows from the point of failure. Each pipeline step is implemented as a callable function within the \texttt{AVICA} library, which \texttt{ALFRD} invokes in sequence according to the configured workflow. To manage these runs, \texttt{ALFRD} maintains a registry of all datasets, tracks their status at each processing stage, and generates an active log. This registry can be provided in common tabular formats, such as CSV files or spreadsheets, with each row corresponding to a single dataset.

A key feature of \texttt{ALFRD} is the ability to attach an external spreadsheet, such as a Google Sheet, to the pipeline workflow. This spreadsheet serves as a human-readable progress monitor, recording the processing status of each dataset as it advances through the calibration stages. Entries are updated automatically as workflows are dispatched and completed, providing a real-time overview of the pipeline progress - including the number of datasets successfully calibrated, those currently being processed, and those that have encountered failures requiring attention. This makes it straightforward to monitor the state of a large batch run without inspecting individual log files. Upon completion of each dataset, \texttt{ALFRD} collects the resulting output and diagnostic products and writes the outcomes back to the progress spreadsheet. 

\section{Data calibration in VLBI}\label{sec:calibrationvlbi}

 The calibration of interferometric data is naturally described within the Radio Interferometry Measurement Equation (RIME) framework \citep{Hamaker1996, Smirnov2011p1}, in which each physical effect along the signal path is represented as a Jones matrix acting on the true sky signal. In the \texttt{CASA} implementation, the cumulative corruption $\mathcal{J}_{ij}$ along the signal path recorded at antennas $i$ and $j$ is expressed as:
\begin{equation}
    \mathbf{V}_{ij}^{\rm obs} = \mathcal{J}_{ij} \, 
    \mathbf{V}_{ij}^{\rm true}
\end{equation}
where $\mathbf{V}_{ij}^{\rm obs}$ and $\mathbf{V}_{ij}^{\rm true}$ are the observed and true complex visibi{\-}lities obtained by correlating the signals recorded at antennas $i$ and $j$. The term $\mathcal{J}_{ij}$ encapsulates a hierarchy of direction-independent effects including amplitude errors from sampler statistics and system temperature variations, bandpass distortions across the observed frequency range, and residual phase, delay, and delay-rate offsets arising from clock instabilities and atmospheric fluctuations \citep{JanssenSoftware2022, Thompson2017}. A key consequence of the RIME formulation is that calibration can be decomposed into a sequence of independent steps, each targeting a specific Jones term using observations of a bright, point-like calibrator source. For connected-element interferometers, many of these terms are either stable or slowly varying and can be solved using standard self-calibration techniques. In VLBI, however, the independent recording of signals at widely separated stations introduces rapidly varying phase and delay terms that dominate the corruption budget, requiring a dedicated fringe-fitting step.

\subsection*{Fringe fitting in \texttt{CASA}}\label{sec:fringefitoverview}

Fringe fitting solves for the residual group delay, delay rate, dispersive delay, and phase on each antenna effectively removing contributions of clock offsets and atmospheric fluctuations between stations that are not absorbed by the correlator model. The \textit{fringefit} task implemented in \texttt{CASA} \citep{bemmelCASAonfringe2022} follows the mathematical framework of \citet{schwabandcotton1983}, which also underlies the \texttt{FRING} task in \texttt{AIPS}.

Residual instrumental and propagation effects cause the observed visibility phase $\tilde{\phi}_{ij}$ on each baseline $i$--$j$ to deviate from the true phase $\phi_{ij}$ according to:
\begin{equation}
  \tilde{\phi}(t,\nu) = \phi_{ij}(t,\nu)
              + (\phi_{i0}-\phi_{j0})
              + \Delta \nu(\tau_i - \tau_j)
              + \Delta t(\dot\tau_i - \dot\tau_j)
              + \phi_\mathrm{disp}
  \label{eq:phasemodel}
\end{equation}

where $\Delta t = t - t_\mathrm{ref}$ and $\Delta \nu = \nu - \nu_\mathrm{ref}$ are the offset from the reference time and frequency, $\phi_0$ is constant phase, $\tau$ is the residual group delay; $\dot{\tau}$ is the delay rate, and $\phi_\mathrm{disp}$ is a dispersive delay term capturing ionospheric contributions proportional to $\nu^{-1}$ for dispersive delay \citep[see][]{Small2022, martividal2010}.
The \textit{fringefit} task solves for all four parameters per solution interval; each parameter can individually be included or excluded via the \textit{paramactive} argument. By default, $\tau$ and $\dot{\tau}$ are solved for and the dispersive term is not. The constant phase $\phi_0$ is always included in the solve. The \textit{fringefit} with global solve is done by first performing a fringe detection using the Fast Fourier Transform (FFT) on each baseline to a reference antenna on a specified solution interval. This FFT is treated as an initial guess to do a global least-squares solve on an assumed model. By default, a point source model is assumed. Alternatively, the task uses the model provided if a model data column is present in the MS file. The grid for the FFT search is constrained by defining specific delay and rate windows. The \textit{fringefit} task supports single-band mode or multi-band mode, the latter combines solutions from coherent spectral windows using \textit{concatspws=True} and \textit{combine='spw'} parameters. When available, RR and LL correlations can be further combined to improve fringe detection, using the \textit{corrcomb} parameter (\textit{'parallel'}, \textit{'stokes'}, or \textit{'none'}). Combining data across spectral windows or correlations for the fringe-fit should only be performed after all instrumental calibration terms have been corrected.

\section{Pipeline workflow}\label{sec:workflow}

\begin{figure*}[htbp]
    \centering
    \includegraphics[width=\linewidth, trim=0cm 1.05cm 0cm 1.05cm, clip]{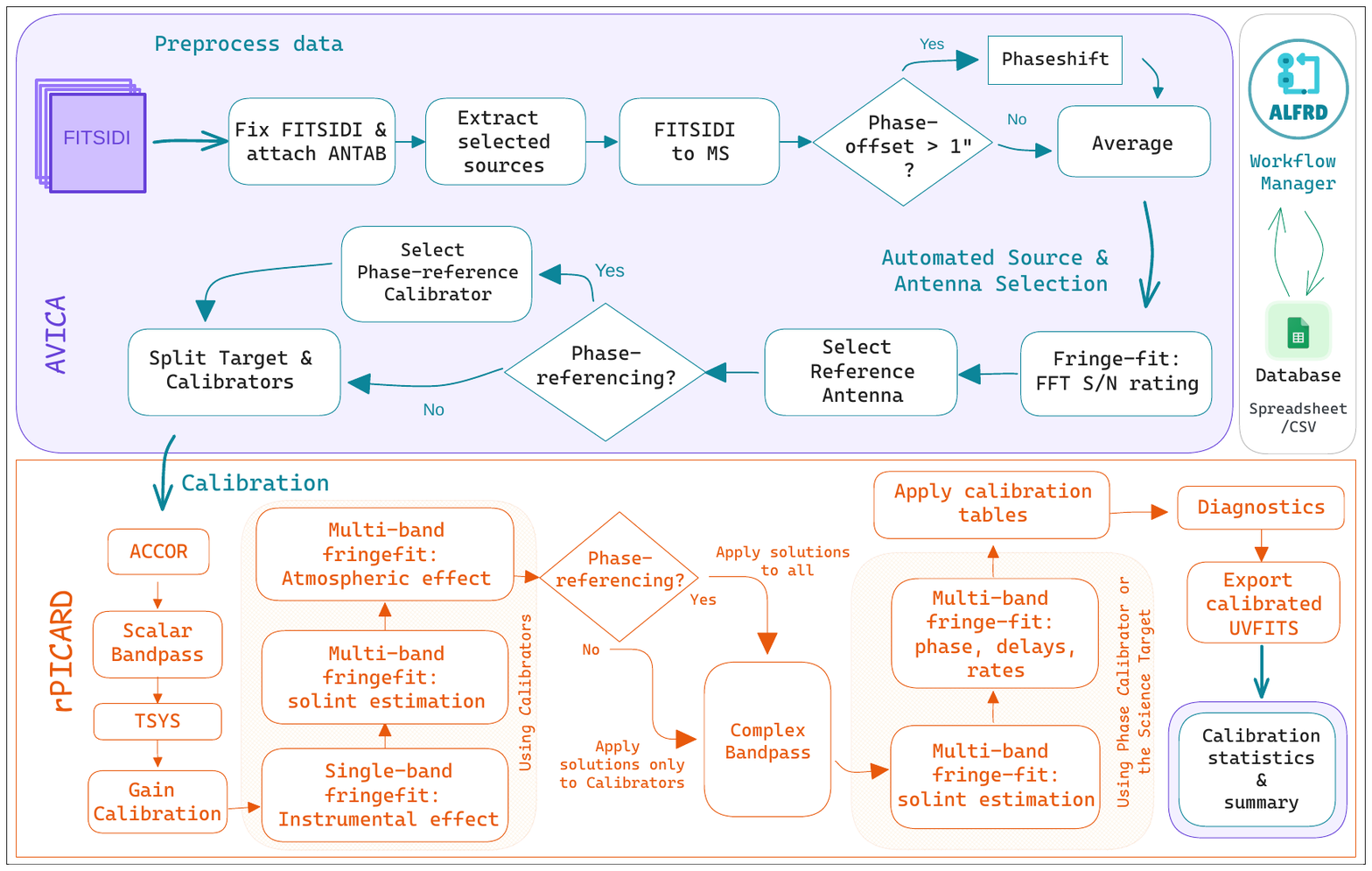}
    \caption{Complete \texttt{AVICA} pipeline workflow. The preprocessing stage (top, purple) operates on the raw FITS-IDI and MS data, handling source extraction, format corrections, phase center shift, averaging, and the automated selection of calibrators and reference antennas. The calibration stage (bottom, orange) follows the \texttt{rPICARD} scheme, executing in sequence amplitude calibration, single-band fringe-fitting for instrumental effects, multi-band fringe-fitting to correct rapidly varying atmospheric phases, complex bandpass calibration, and finally phase calibration via multi-band fringe-fitting, with solutions derived from the phase-reference calibrator in phase-referencing mode, or directly from the science target otherwise. The pipeline concludes with the generation of calibration statistics and a diagnostic summary (bottom right). The end-to-end execution, indicated by the teal arrows, is orchestrated by the \texttt{ALFRD} workflow manager, which tracks progress via a central database. Decision diamonds indicate conditional branches.}
    \label{fig:pipelineworkflow}
\end{figure*}

The pipeline takes the raw input data for calibration in the FITS-IDI format. Depending on the specific data requirements, the \texttt{AVICA} pipeline (Fig.~\ref{fig:pipelineworkflow}) performs preprocessing in the FITS-IDI before and in the MS data after loading the data in \texttt{CASA}.

\subsection{Processing data: pre-loading}\label{sec:preprocessFITSIDI}

The raw data retrieved from the NRAO archive span formats that have evolved between 1995 and 2023, and an initial preprocessing step is required to correct formatting inconsistencies before calibration can proceed. These include, among others, binary data erroneously present in ASCII table extensions of the FITS-IDI file, and repeated antenna entries where the same antenna appears under multiple IDs while the data references only one, in which case the unused duplicate is removed; a complete list of handled inconsistencies is provided in the documentation\cref{avicadocumentation}.
These inconsistencies are handled automatically by the \textit{fitsidi\_check} function implemented within \texttt{AVICA}. The standard for the FITS-IDI is extensively defined in the \texttt{AIPS Memo 114}\footnote{\label{memo114}\url{https://library.nrao.edu/public/memos/aips/memos/AIPSM_114.pdf}}.

Several observations in the SMILE sample of $\sim5000$ sources have very large file sizes of more than $100\,\rm{GB}$ containing hundreds of sources in a single file. In our tests, the loading of these datasets absorbed most of the time in the pipeline. To reduce the execution time, only the required calibrator and target sources are extracted from the FITS-IDI file prior to loading. This extraction is triggered when the number of sources in the file exceeds 50 and the file size is above $100\,\rm{GB}$. Calibrator sources are identified by cross-matching the sources present in the file against the Radio Fundamental Catalogue (RFC) \citep{2025Petrov&Kovalev}, with the VLBA calibrator list serving as a fallback, retaining only sources whose coordinates fall within 100 milliarcseconds of a cataloged calibrator entry. The top $10$ matched sources, ranked by flux density within the respective frequency bands, are retained alongside the target source. All thresholds and parameters described above are configurable by the user. Since no package presently supports direct source extraction from the FITS-IDI format, this extraction is handled by \textit{fitsidiutil}, a preprocessing module supplied as part of the \texttt{AVICA} package. Where necessary, \textit{fitsidiutil} also separates the data by frequency ID, producing individual FITS-IDI files per frequency ID, as \texttt{CASA} is currently limited to reading FITS-IDI files containing only a single frequency ID. The individually split files are then treated as separate observations. Some observations predating the \texttt{DiFX} correlator \citep{deller2007} have visibility data distributed across multiple files, which must be loaded together. 

Additionally, if the gain curve or system temperature data is missing from the FITS-IDI file, the pipeline automatically downloads them from the NRAO archive. These data are stored in the archive as system temperature and gain curve tables, from which the pipeline generates the  flux calibration table in the ANTAB\footnote{\url{https://avica.readthedocs.io/en/latest/misc/antab.html}} format. The ANTAB file is then appended to the FITS-IDI files using the \texttt{casa-vlbi}\footnote{\url{https://github.com/jive-vlbi/casa-vlbi}} Python package. Finally, the FITS-IDI files are loaded into the MS file using the \texttt{CASA} task \textit{importfitsidi}.

\subsection{Processing data: post-loading}\label{sec:preprocessMS}

The native data format of \texttt{CASA} is the MS file, which is built on the Casacore Table Data System \citep{DiepenCasaTableMS2015} - a relational-like data model capable of storing large arrays. \texttt{CASA} provides libraries for retrieving metadata and calibrating the visibility data stored in the MS format.

After loading the data, sources are examined for a possible offset of the correlation phase center from the true source coordinates. The true coordinates can be supplied in two ways: as a CSV file listing source names and their corresponding coordinates for batch processing or as a single coordinate entry for an individual source, both specified in the input configuration file. If an offset of more than one arcsecond is found between the phase center and the supplied coordinates, the \texttt{CASA} task \textit{phaseshift} is executed to re-center the visibility data accordingly.

If multiple frequencies are present, the MS data are split into these separate frequency bands, each band possibly requiring an independent calibration workflow with its own set of calibrators and reference antennas.
Each band is subsequently averaged in time and frequency, adopting a time bin of $2\,$s and a channel width of $500\,$ kHz, provided the native resolution is finer than $1\,$s and $500\,$ kHz, as discussed in \citet{VIPCALs2025}. These averaging parameters represent an empirical balance between reducing thermal noise and keeping time and frequency smearing effects within acceptable limits, for data typically observed at high time and frequency resolution for astrometric purposes. The resulting data are also significantly reduced in size, thereby decreasing the computational time of the subsequent calibration steps. Additionally, any antennas not required for the calibration are removed at this stage. This is fully user-configurable, allowing the pipeline to be adapted to different array combinations. For example, in observations where VLBA antennas are mixed with stations from EVN, only the VLBA antennas are retained by default for the SMILE project, as this simplifies both calibration and imaging, and ensures a consistent data sample for comparison with the \texttt{AIPS}-based \texttt{VIPCALs} pipeline. Mixed-array data can be processed by adjusting this setting accordingly. From this point on, each band is treated as an independent observation and follows its own separate calibration branch.

\subsection{Selecting reference antenna and calibrators}

The reference antenna serves as the phase and delay reference for all other antennas throughout the observation, any gaps in its coverage directly propagate as gaps in the calibration solutions for the entire array, making its reliability critical. To mitigate this, a ranked list of reference antennas should be provided, allowing \texttt{rPICARD} to re-reference solutions to a substitute antenna as needed, ensuring continuity in the fringe-fit solutions without gaps or discontinuities. Thus, an ideal reference antenna is one that is available for the longest duration of the observation, achieves a high median FFT S/N across all baselines, and is located near the geometric center to minimize average baseline lengths, thereby maximizing the likelihood of a detectable fringe on every reference baseline at the FFT stage. By convention, the group delay, delay rate, and phase (given in Eq. \ref{eq:phasemodel}) of the reference antenna are set to zero, and the parameters for all other antennas are solved relative to this reference, since the antenna parameters appear only as differences between antenna pairs and cannot be determined in an absolute sense.

For each band, a fringe fit is performed on all available antennas across all available scans, using a solution interval spanning the full scan duration. This fringe fit is performed using the FFT-based solver without the global least-squares algorithm so that the resulting S/N reflects the strength of the fringe detection directly in the data. The global least-squares S/N, by contrast, depends on the assumed source model and the residuals of an iterative fit \citep{schwabandcotton1983}, making it less suited for an unbiased assessment of fringe detectability at this stage. The FFT-based S/N therefore provides a more direct, model-independent measure of the per-antenna detection strength, making it better suited for rating the quality of candidate reference antennas and calibrators. For observations that have more than 50 scans, a subset is selected based on scan duration and antenna availability, with priority given to scans central to the observation. This selection improves the time spent during the pipeline run on the S/N rating step for observations containing hundreds of scans.

The reference antenna is selected based on a combination of the FFT S/N and its proximity to the centroid of the array. The centroid is computed from all available antennas in the given band, and the antenna that best balances a high median S/N across all baselines with a small distance from the array centroid is chosen. The selected reference antenna must also satisfy the two following conditions: it must remain unflagged for the majority of the calibrator scans, and it must be available throughout the duration of the science target scans. For cases where no single antenna satisfies these conditions across all scans, a ranked combination of antennas is chosen as the reference, used together with the \textit{refantmode=`flex'} parameter in the \textit{fringefit} task, which iterates through the prioritized list and assigns the most suitable available antenna at each calibration step. Finally, solutions are re-referenced to a common reference antenna to preserve phase continuity across the observation.

A bright calibrator source typically produces significantly higher FFT S/N than the comparatively faint science targets. The pipeline defaults to selecting the top $5$ sources as calibrators, ranked by their median FFT S/N across all baselines and scans, subject to the condition that each selected source maintains maximum antenna availability across its associated scans. The number of calibrators is configurable by the user.

Phase-referencing mode is an optional, user-configurable feature of the pipeline, disabled by default. When enabled, the pipeline automatically identifies phase-referencing observations by examining the scan sequence. \texttt{AVICA} checks the complete scan list and, if an interleaved calibrator scan is found within 15 degrees of the science target, selects it as the phase-reference calibrator. If the median FFT S/N of the science target, evaluated across all scans during the S/N rating step, exceeds a user-configurable threshold, set to 6 by default, the pipeline bypasses the phase-reference calibrator selection and performs fringe fitting directly on the science target instead. The angular separation threshold is also user-configurable.

\subsection{Calibration}\label{sec:calibration}

The calibration process addresses errors stemming from signal propagation as well as hardware and correlator imperfections, following the method described extensively in \citet{Janssen2019}. Here we provide a brief outline of the steps as implemented within \texttt{AVICA}.

The full \texttt{rPICARD} calibration scheme is executed for each dataset dispatched by \texttt{ALFRD} (Fig.~\ref{fig:pipelineworkflow}). First, sampler corrections are applied using the autocorrelations to account for amplitude errors introduced by digitization in the correlation process. A scalar bandpass correction is then performed to fix the shape of the passband in amplitude as a function of frequency using auto-correlations, followed by amplitude calibration from system equivalent flux densities using the system temperature and gain curve tables. Instrumental errors in phase and delay are subsequently corrected by running the \textit{fringefit} task on the brightest calibrator source, making the spectral windows within each band coherent. A multi-band fringe-fit is then performed to obtain solutions for the rapidly varying atmospheric phases, corrections for which are applied only to the calibrators so that the subsequent bandpass solution is not corrupted by residual atmospheric phase fluctuations. The complex bandpass calibration then accounts for phase variations as a function of frequency for each spectral window using cross-correlations, with an additional amplitude correction applied if the scalar bandpass correction was not performed. Following that, phase calibration is performed using a multi-band fringe fit. In phase-referencing mode, solutions are derived from the phase-reference calibrator and transferred to the science target. Otherwise, the fringe fit is performed directly on the science target.

Before each fringe-fitting step, the coherence solution interval is estimated by iterating through a range of candidate intervals, which can be specified by the user or determined automatically following the procedure described in \citet{Janssen2019}. Scans for which no valid fringe solution is found are automatically flagged and excluded from subsequent calibration steps.  All calibration tables are accumulated throughout the run and applied to the data collectively at the end of each run. Diagnostic plots are produced at the end of calibration using \texttt{jplotter}\footnote{\url{https://github.com/haavee/jiveplot.}}, together with a calibration summary.

\subsection{Flagging}\label{sec:flagging}

The SMILE sample represents a diverse range of observations, including those taken for astrometric purposes that may contain very little visibilities per observation. To accommodate this diversity and avoid inadvertently flagging valid data, standard flagging procedures such as quacking to remove data corrupted by antenna slewing at scan boundaries were omitted. The only flagging applied is a configurable seven percent flag to the edge channels of each spectral window to account for the bandpass roll-off effect. Automatic flagging tools for radio frequency interference, such as \texttt{AOFlagger} \citep{Offringa2010}, are compatible with the MS format and could, in principle, be incorporated into the pipeline; the flexible design of \texttt{AVICA} makes such an integration straightforward in future versions. 

\section{Results and Discussion}\label{sec:results}

\subsection{VLBA test sample}\label{sec:testsample}
The \texttt{AVICA} pipeline was tested on a subsample of 1000 sources from the SMILE sample. The dataset comprises nearly three decades of heterogeneous archival VLBA observations. This time range encompasses changes in correlator technology, recording bandwidth, and data format -- from pre-DiFX observations recorded on legacy hardware \citep{Benson1995} to modern wideband datasets correlated with the DiFX correlator software \citep{deller2007}. The sample therefore represents a challenging testbed for fully automated calibration, as no two datasets are guaranteed to share the same frequency configuration, antenna availability, or file structure. Individual FITS-IDI files in the sample range from a few gigabytes to over 100\,GB, with many large files containing hundreds of sources observed in a single scheduling block. The tested sample is described in detail in \citet{VIPCALs2025}, where a calibration of the same sample using the \texttt{VIPCALs} pipeline is also presented.

\texttt{AVICA} successfully calibrated 978 out of 1000 sources, yielding calibrated visibilities in both MS and UVFITS\cref{memo114} formats. The two output formats ensure compatibility across different analysis environments: the MS files allow for continued post-processing within \texttt{CASA}, while the UVFITS files are readily compatible with external radio imaging software such as \texttt{Difmap} \citep{Shepherd1997}. The pipeline processed $1372$ individual band-separated observations in total, spanning the S, C, X, U, and K bands, with the C and X bands comprising the majority of the sample. These frequency bands are ideal for the SMILE project because of the compromise between resolution, sensitivity, and minimal atmospheric effects. 

\begin{figure}
    \centering
    \includegraphics[width=\linewidth]{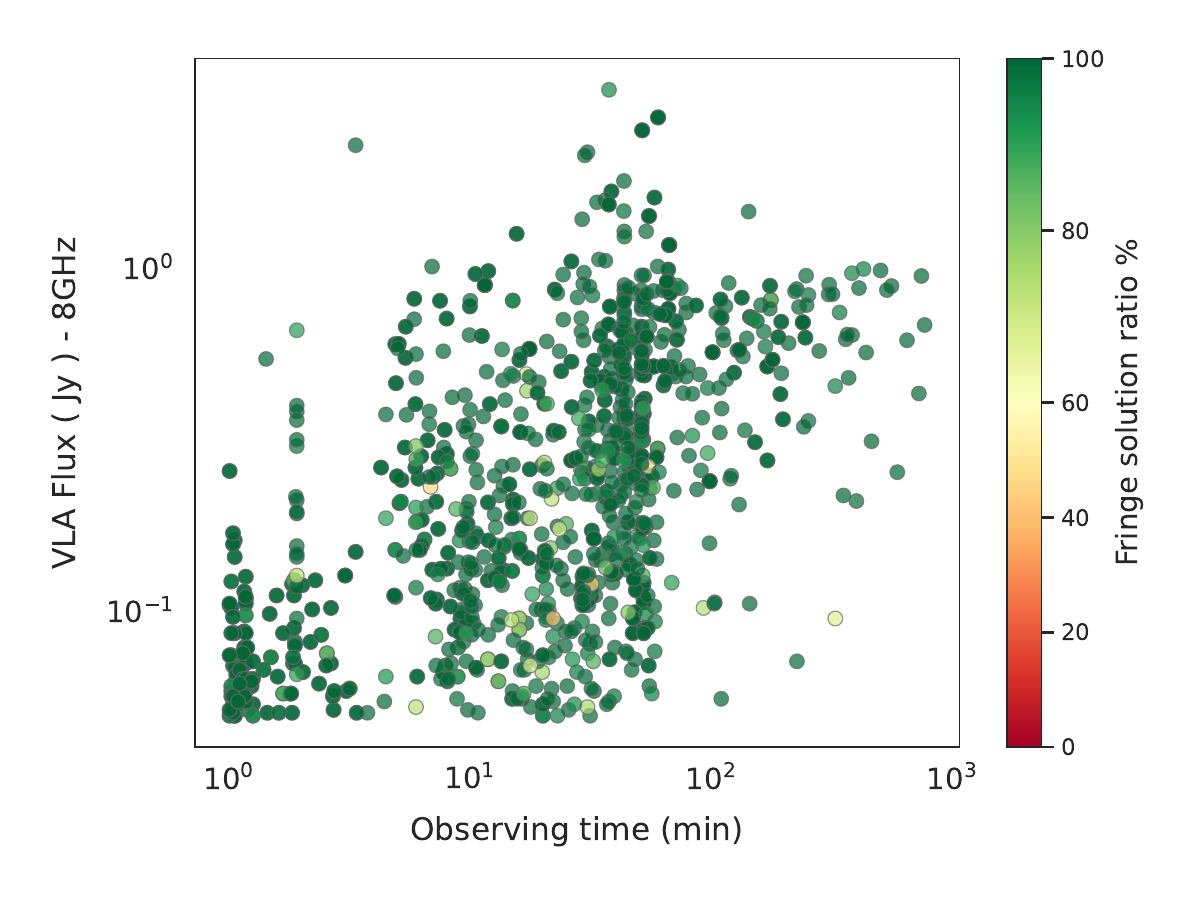}
    \caption{VLA flux density at 8\,GHz versus total observing time for the $1372$ band-separated observations in the test sample, color-coded by the fringe solution ratio, defined as the fraction of fringe-fit solutions exceeding the minimum $\mathrm{S/N}$ detection threshold of $3.2$. 
    }
    \label{fig:fringedistribution}
\end{figure}

\subsection{Calibration result}\label{sec:calibrationresult}
 The $22$ datasets that could not be processed fall into two categories: incomplete FITS-IDI files and missing system temperature data. Of these failures, $4$ datasets failed prior to loading into the MS format due to unreadable input files, while $3$ datasets failed due to the absence of system temperature records in the NRAO archive logs. The remaining 15 failures stemmed from metadata inconsistencies in the system temperature records such as missing headers or typographical errors, which required manual rectification.

 Phase calibration on the science target is performed using the \textit{fringefit} task in \texttt{CASA} with the least-squares global solver, performing multi-band calibration. The quality of the calibration can be assessed through the fringe solution percentage, defined as the percentage of solutions exceeding a minimum S/N threshold divided by the total number of solutions. We adopted a $\mathrm{S/N}$ threshold of $3.2$, for the calibration of the tested data, following the default parameter from \texttt{rPICARD}. The S/N for the \textit{fringefit} task in \texttt{CASA} is, by default, set to $3.0$. 
 Together with the fringe solution percentage, the visibility retention ratio is used to assess calibration quality, defined as the number of unflagged visibilities in the final calibrated target data divided by the total number in the uncalibrated data.
 
 Fig.~\ref{fig:fringedistribution} shows the VLA flux density from the CLASS catalog at 8\,GHz versus total observing time for the $1372$ band-separated observations, color-coded by the fringe solution percentage.
 
\subsection{Comparison with \texttt{VIPCALs}}\label{sec:comparewithVIPCALs}
\begin{figure}[t]
    \centering
    \includegraphics[width=\linewidth]{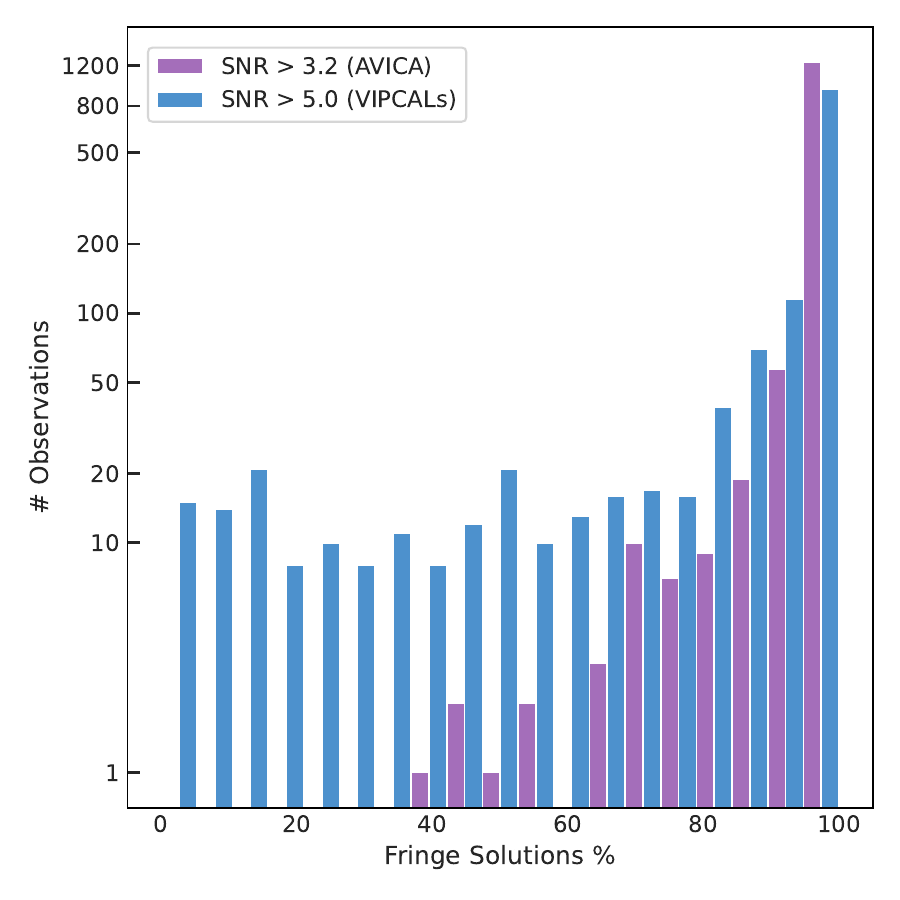}
    \caption{Distribution of fringe solution percentages for the band-separated $1372$ observations. The \texttt{AVICA} pipeline is shown at its default $\mathrm{S/N}$ threshold of $3.2$ (purple), alongside the \texttt{VIPCALs} distribution at $\mathrm{S/N}$ of $5.0$ (blue).}
\label{fig:fringecomparison}
\end{figure}

\begin{figure}[]
    \centering
    \includegraphics[width=\linewidth]{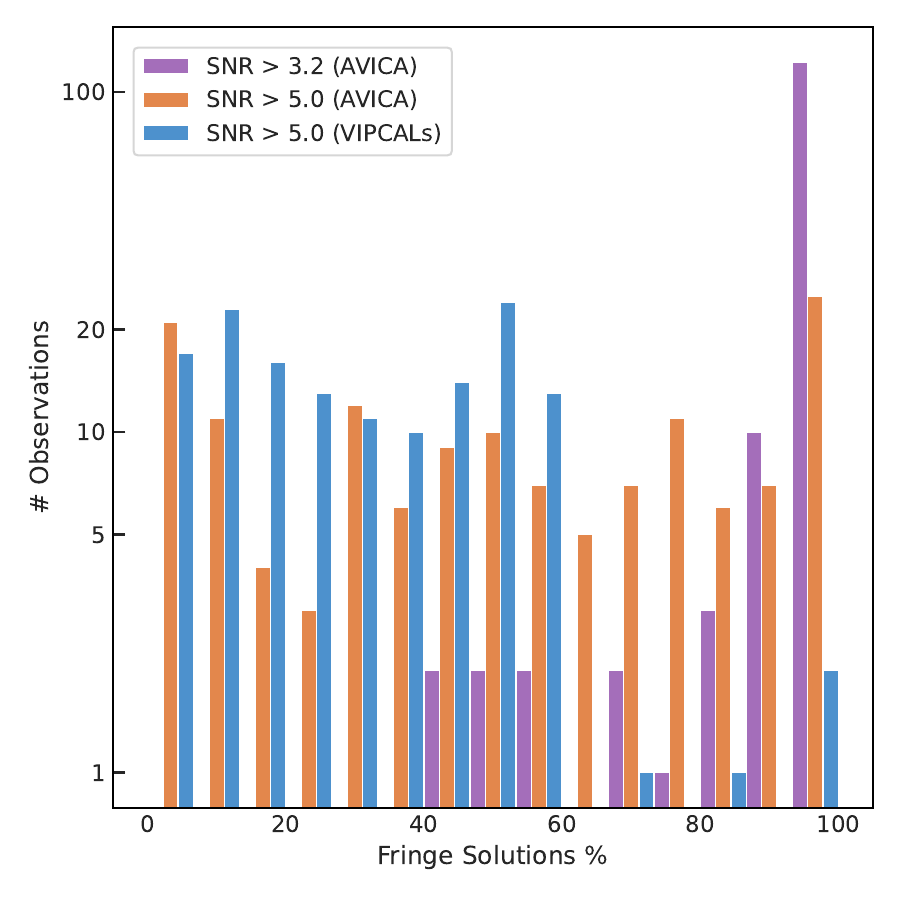}
    \caption{Comparison of fringe solution percentages for a subset of $145$ band-separated observations in which either \texttt{AVICA} or \texttt{VIPCALs} fell below a $60\%$ solution rate at their respective default threshold. The purple bars represent \texttt{AVICA} at its default $\mathrm{S/N}$ threshold of $3.2$, the blue bars \texttt{VIPCALs} at its standard $\mathrm{S/N}$ threshold of $5.0$, and the orange bars \texttt{AVICA} re-run at the matching $\mathrm{S/N}$ threshold of $5.0$.}
\label{fig:fringecomparisonunder60perc}
\end{figure}

\begin{figure*}[htbp]
    \centering
    \includegraphics[width=0.96\linewidth, trim=0cm 1.4cm 0cm 0cm, clip]{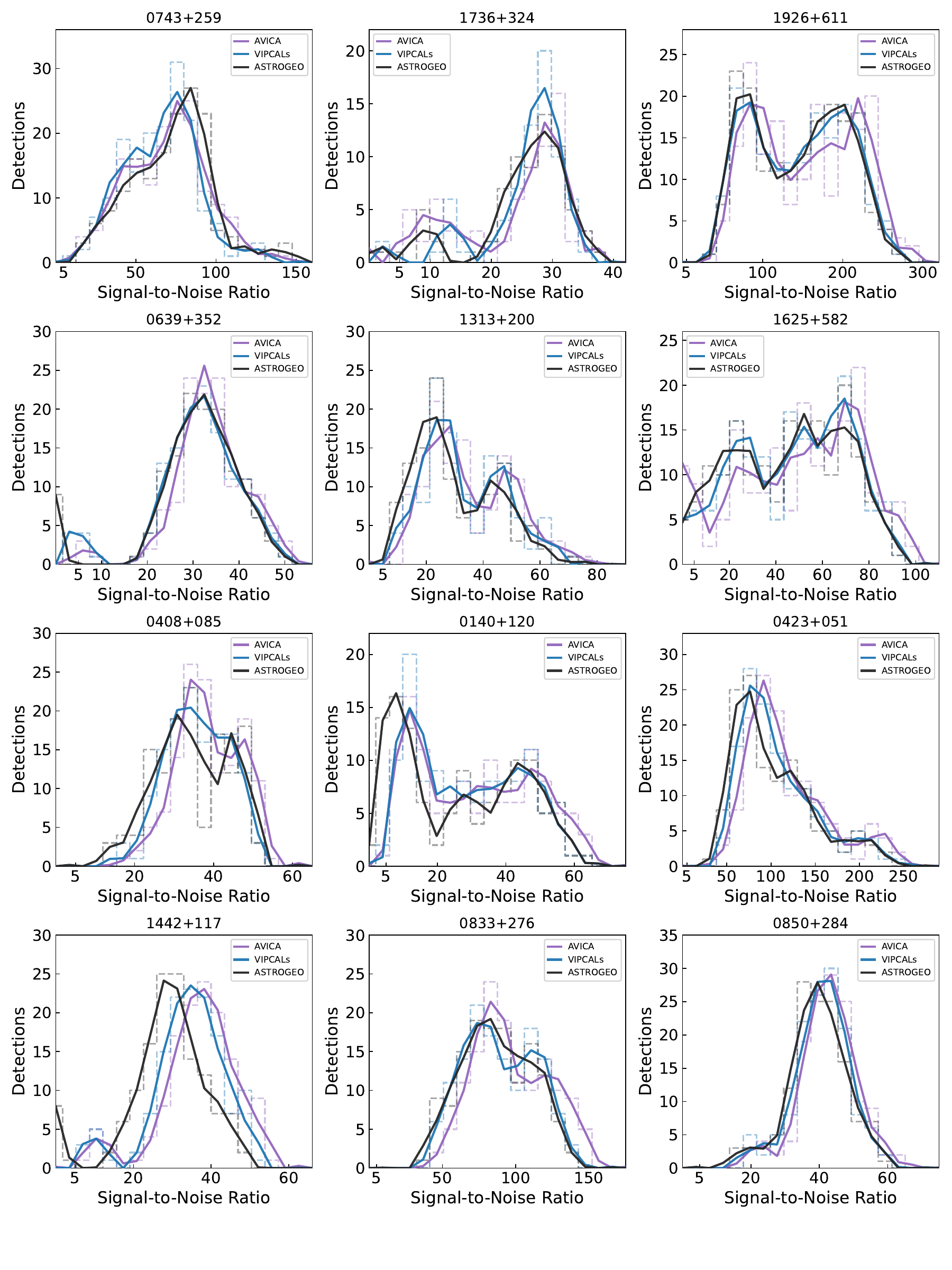}
    \caption{Distributions of $\rm{S/N}$  of the per-scan, per-baseline detections for the 12 \texttt{Astrogeo} X-band VLBA datasets (Table~\ref{table:FluxVSObs}), one panel per source. The three reduction pathways are shown as \texttt{AVICA} (purple), \texttt{VIPCALs} (blue), and the manually calibrated \texttt{Astrogeo} reference (black). Faint dashed lines show the raw histograms (20 bins). Solid curves show the same counts after a Savitzky-Golay filter is applied for visual clarity. The vertical axis gives the number of detections per bin.}
\label{fig:plot_sn_dist}
\end{figure*}

\begin{figure*}[]
    \centering
    \includegraphics[width=0.96\linewidth, trim=0cm 1.4cm 0cm 0cm, clip]{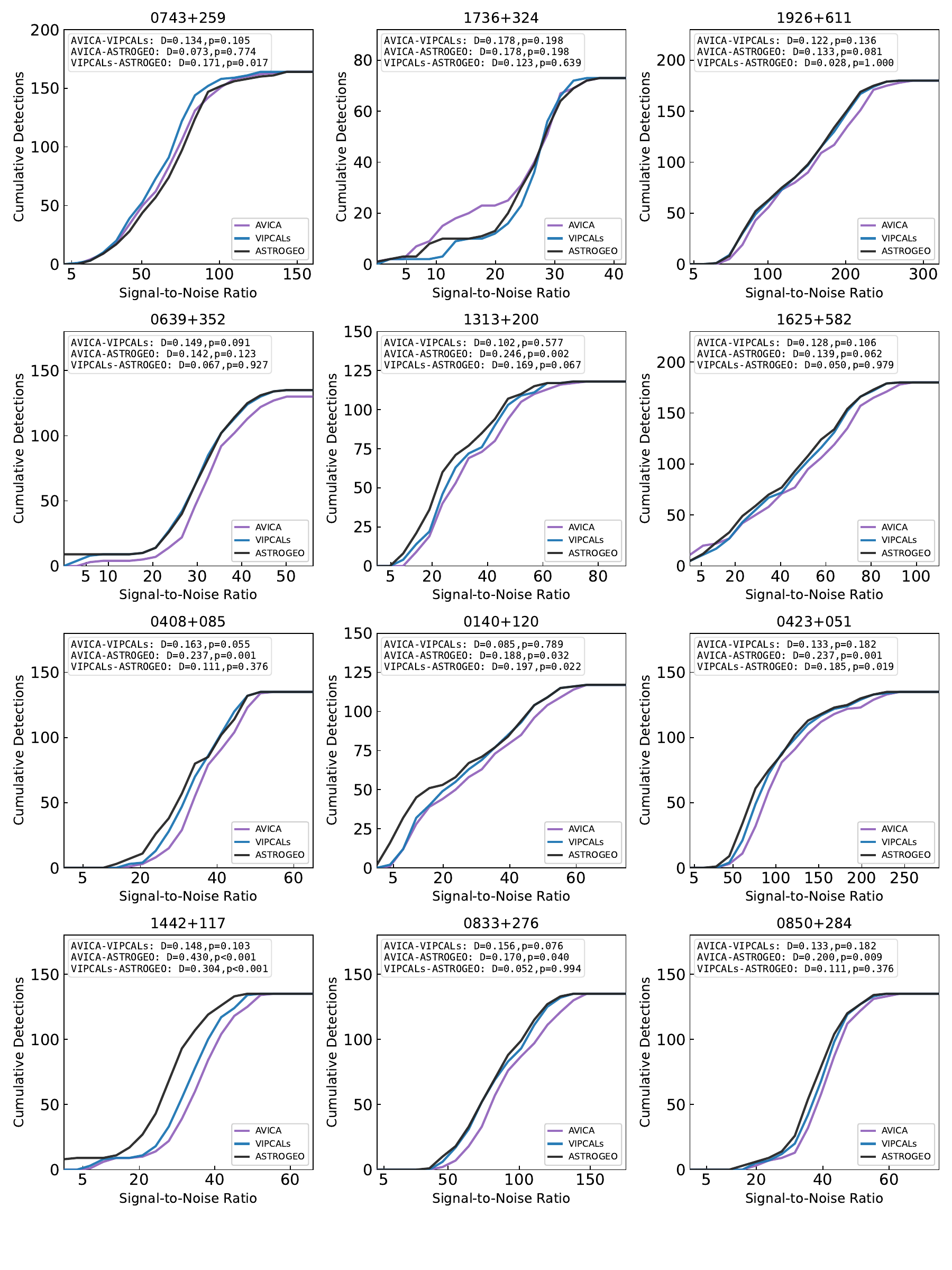}
    \caption{Cumulative distributions of the $\rm{S/N}$ for the same 12 datasets as in Fig.~\ref{fig:plot_sn_dist}, with one panel per source. Curves show \texttt{AVICA} (purple), \texttt{VIPCALs} (blue), and the manually calibrated \texttt{Astrogeo} reference (black). The annotation in each panel reports the two-sample Kolmogorov-Smirnov statistic $D$, the maximum vertical separation between two CDFs, and the associated $p$-value for each pair of pipelines.}
\label{fig:cdf_plot}
\end{figure*}

Fig.~\ref{fig:fringecomparison} illustrates the distribution of fringe solution percentages for the sample of $1372$ band-separated observations. The \texttt{AVICA} results at the default fringe-fitting $\rm{S/N}$ threshold of $3.2$ (purple) are compared with those of \texttt{VIPCALs} at its standard $\mathrm{S/N}$ threshold of $5.0$ (blue). Both pipelines show broad agreement at high solution percentages, with the majority of observations exceeding 80\%. However, at the lower end of the distribution, the stricter \texttt{VIPCALs} threshold results in a higher number of observations in low-percentage bins compared to the default \texttt{AVICA} run.

To investigate this further, we identified 145 band-separated observations where the fringe solution percentage fell below $60\%$ for either \texttt{VIPCALs} or \texttt{AVICA} at their respective default thresholds. These observations were re-calibrated at a stricter $\rm{S/N}$ threshold of $5.0$ and are shown in Fig.~\ref{fig:fringecomparisonunder60perc}. At the default \texttt{AVICA} $\rm{S/N}$ threshold of $3.2$, 139 of these observations yield solution percentages above $80\%$ (purple), with only 6 falling below $60\%$. When the $\rm{S/N}$ threshold is raised to $5.0$, the distribution broadens significantly: 37 observations drop below 20\%, 48 remain in the 20--60\% range, and the remaining 60 recover to a mean solution percentage of $92.29\%$. The recovery of this last group, comprising $41.4\%$ of the re-calibrated sample, confirms that a significant fraction of the band-separated observations at $\rm{S/N}$ threshold of $3.2$ correspond to genuinely detectable fringes rather than noise.
The $37$ observations that fall below $20\%$ under the stricter threshold are consistent with sensitivity-limited detections, yielding poor fringe solutions regardless of the pipeline used. These observations represent marginal solutions in noisy data that are typically refined or flagged during the subsequent imaging stage.

The close agreement between the results from the \texttt{AVICA} and \texttt{VIPCALs} distributions at $\rm{S/N}$ threshold of $5.0$ confirms that the discrepancies at the low end of the distribution are driven by the choice of S/N threshold rather than by differences in calibration quality between the two pipelines.

\begin{table}
\caption{Properties of the 12 \texttt{Astrogeo} X-band VLBA datasets used for the comparison: Source name (B1950), Project code, on-source observation length (s), and total flux density (Jy) taken from the \texttt{Astrogeo} archive.}
\label{table:FluxVSObs}
\centering
\begin{tabular}{l c c c}
\hline\hline
Source & Project & Obs.\ length & Flux \\
(B1950)   &    & (s) & (Jy) \\
\hline
0743+259 & BG219C & 218  & $0.341$ \\
1736+324 & RDV88 & 2412 & $0.129$ \\
1926+611 & UF001D & 274  & $0.733$ \\
0639+352 & UF001H & 286  & $0.113$ \\
1313+200 & UF001J & 436  & $0.134$ \\
1625+582 & UF001J & 480  & $0.174$ \\
0408+085 & UF001K & 288  & $0.113$ \\
0140+120 & UF001L & 346  & $0.125$ \\
0423+051 & UF001M & 434  & $0.424$ \\
1442+117 & UF001O & 476  & $0.089$ \\
0833+276 & UF001P & 168  & $0.414$ \\
0850+284 & UF001P & 228  & $0.136$ \\
\hline
\end{tabular}
\end{table}

\subsection{Comparison with manual calibration}
\begin{table*}[ht]
\caption{Two-sample Kolmogorov--Smirnov statistics $D$ and $p$-values comparing the S/N distributions of the three reduction pathways for the 12 \texttt{Astrogeo} datasets.}
\label{table:KStest}
\centering
\begin{tabular}{l cc cc cc r}
\hline\hline
 & \multicolumn{2}{c}{AVICA--VIPCALs} & \multicolumn{2}{c}{AVICA--Astrogeo} & \multicolumn{2}{c}{VIPCALs--Astrogeo} & Detections \\
Source & $D$ & $p$ & $D$ & $p$ & $D$ & $p$ & $N_{\rm det}$ \\
\hline
0743+259 & $0.134$ & $0.105$ & $0.073$ & $0.774$ & $0.171$ & $0.017$ & 164 \\
1736+324 & $0.178$ & $0.198$ & $0.178$ & $0.198$ & $0.123$ & $0.639$ & 73 \\
1926+611 & $0.122$ & $0.136$ & $0.133$ & $0.081$ & $0.028$ & $1.000$ & 180 \\
0639+352 & $0.149$ & $0.091$ & $0.142$ & $0.123$ & $0.067$ & $0.927$ & $^{*}135$ \\
1313+200 & $0.102$ & $0.577$ & $0.246$ & $0.002$ & $0.169$ & $0.067$ & 118 \\
1625+582 & $0.128$ & $0.106$ & $0.139$ & $0.062$ & $0.050$ & $0.979$ & 180 \\
0408+085 & $0.163$ & $0.055$ & $0.237$ & $0.001$ & $0.111$ & $0.376$ & 135 \\
0140+120 & $0.085$ & $0.789$ & $0.188$ & $0.032$ & $0.197$ & $0.022$ & 117 \\
0423+051 & $0.133$ & $0.182$ & $0.237$ & $0.001$ & $0.185$ & $0.019$ & 135 \\
1442+117 & $0.148$ & $0.103$ & $0.430$ & $<0.001$ & $0.304$ & $<0.001$ & 135 \\
0833+276 & $0.156$ & $0.076$ & $0.170$ & $0.040$ & $0.052$ & $0.994$ & 135 \\
0850+284 & $0.133$ & $0.182$ & $0.200$ & $0.009$ & $0.111$ & $0.376$ & 135 \\
\hline
\end{tabular}
\tablefoot{The Detections column ($N_{\rm det}$) gives the total number of detections per source, identical across the three pathways except where marked.\\
\tablefoottext{*}{\texttt{AVICA} has 130 detections for \texttt{0639+352} as it flags the \texttt{MK} station.}}
\end{table*}

Manual calibration of VLBI data performed by experienced observers is conventionally regarded as the benchmark for amplitude and phase calibration. We therefore test the calibration performance of \texttt{AVICA} and \texttt{VIPCALs} against a set of manually calibrated data from the \texttt{Astrogeo} VLBI FITS image database \citep{2025Petrov&Kovalev}. From the \texttt{Astrogeo} archive\footnote{\url{https://astrogeo.org/}} we selected 12 calibrated datasets at 8\,GHz, all of which were calibrated manually in \texttt{AIPS} by a single PI. We restrict the comparison to X-band data because this, together with C-band data, will constitute the bulk of the SMILE sample. The datasets were chosen to represent the variety of datasets included in SMILE: total flux densities range from $89\,\mathrm{mJy}$ to $733\,\mathrm{mJy}$ and on-source observation lengths from $2.8$ to $40.2\,\mathrm{minutes}$ (Table~\ref{table:FluxVSObs}), sampling both faint and bright sources and both short and long observing times. The corresponding raw VLBA data were then retrieved from the NRAO archive and calibrated independently with \texttt{AVICA} and \texttt{VIPCALs} at their default settings, giving three reduction pathways for each source: the two automated pipelines and the manual \texttt{Astrogeo} reference.

Each calibrated dataset was first averaged to a 10\,s time bin using \texttt{Difmap}. Rather than adopting the nominal correlator weights, we let \texttt{Difmap} recalculate the visibility weights from the scatter of the data within each averaging bin. The resulting uncertainties reflect the actual noise in each dataset, placing the three reduction pathways on an equal footing.

Visibilities were coherently averaged across each scan and over all spectral windows, making the scan-baseline pair the unit of detection. For each scan-baseline pair we then computed the detection S/N in \texttt{eht-imaging} \citep{Chael2018} as the ratio of the RR correlation amplitude to its uncertainty $\sigma_{\mathrm{RR}}$, the latter following directly from the recalculated weights. Coherent averaging over such long intervals can in principle introduce decoherence if the visibilities vary within a scan. This effect is negligible for our sample, as the sources are all compact RFC calibrators whose \texttt{Astrogeo} images are core-dominated. The RR parallel hand was adopted because it was the only correlation present in all observations, providing a uniform basis for comparison across the three reduction pathways. The total number of detections for a given observation is therefore the sum of the unflagged baselines over all scans. This total is identical for the three reduction pathways on all datasets, with the single exception of \texttt{0639+352}, for which \texttt{AVICA} flags the station Mauna Kea (\texttt{MK}) because the corresponding baseline solutions fell below the S/N threshold during calibration, slightly reducing its detection count relative to \texttt{VIPCALs} and \texttt{Astrogeo}.
 
The $\rm{S/N}$ distributions in Fig.~\ref{fig:plot_sn_dist} are plotted as histograms with 20 bins, with a Savitzky-Golay filter \citep{Savitzky&Golay1964} applied to the binned counts to produce the overlaid smoothed curves for visual clarity. To quantify the agreement, we apply the two-sample Kolmogorov-Smirnov (KS) test to each pair of distributions using the Python function \texttt{ks\_2samp} in \texttt{SciPy} \citep{Virtanen2020} library. The KS test is performed directly on the unbinned per-detection S/N values, independently of the smoothed curves shown for visual clarity. The test assesses the null hypothesis that the two samples are drawn from the same underlying distribution. The KS statistic $D$ measures the maximum vertical separation between two CDFs, serving as the effect size that quantifies how large a difference actually is, and thus provides a single-number summary of the differences visible in Fig.~\ref{fig:cdf_plot}.

The KS statistics are summarized in Table~\ref{table:KStest}. \texttt{AVICA} and \texttt{VIPCALs} are in close mutual agreement: $p>0.05$ for every source, with $D\leq0.18$ across the sample. The small $p$-values seen for several \texttt{AVICA}--\texttt{Astrogeo} pairs reflect the KS test detecting a slight but consistent tendency of \texttt{AVICA} toward higher S/N detections relative to the manual benchmark, an offset not seen to the same degree between \texttt{VIPCALs} and \texttt{Astrogeo}. We note, however, that with $N_{\rm det}\sim70$--$180$ detections per source, the KS test can be sensitive enough to flag even small distributional shifts as statistically significant, so a low $p$-value alone does not imply a practically meaningful difference. The values of $D$ remain small throughout and are the more relevant indicator of agreement. Overall, the three reduction pathways agree within $D\lesssim0.25$ in every case except \texttt{1442+117}, where the effect size reaches $D=0.43$. This outlier arises from marginal detections on baselines to the Hancock (\texttt{HN}) station in a single scan of the \texttt{Astrogeo} observation. A separate peculiarity concerns \texttt{1736+324}, whose total detection count is unusually low: the observation comprises five long ($\sim8$ minute) scans with only 21, 3, 3, 36, and 10 baselines, respectively, giving the substantially lower detection count seen in Figs.~\ref{fig:plot_sn_dist} and~\ref{fig:cdf_plot} despite the long total observation time listed in Table~\ref{table:FluxVSObs}. At this reduced detection count, the offset between AVICA and the other two pathways corresponds to a difference of only 13 detections at the point of maximum divergence.

\begin{figure}[]
    \centering
    \includegraphics[width=0.9\linewidth]{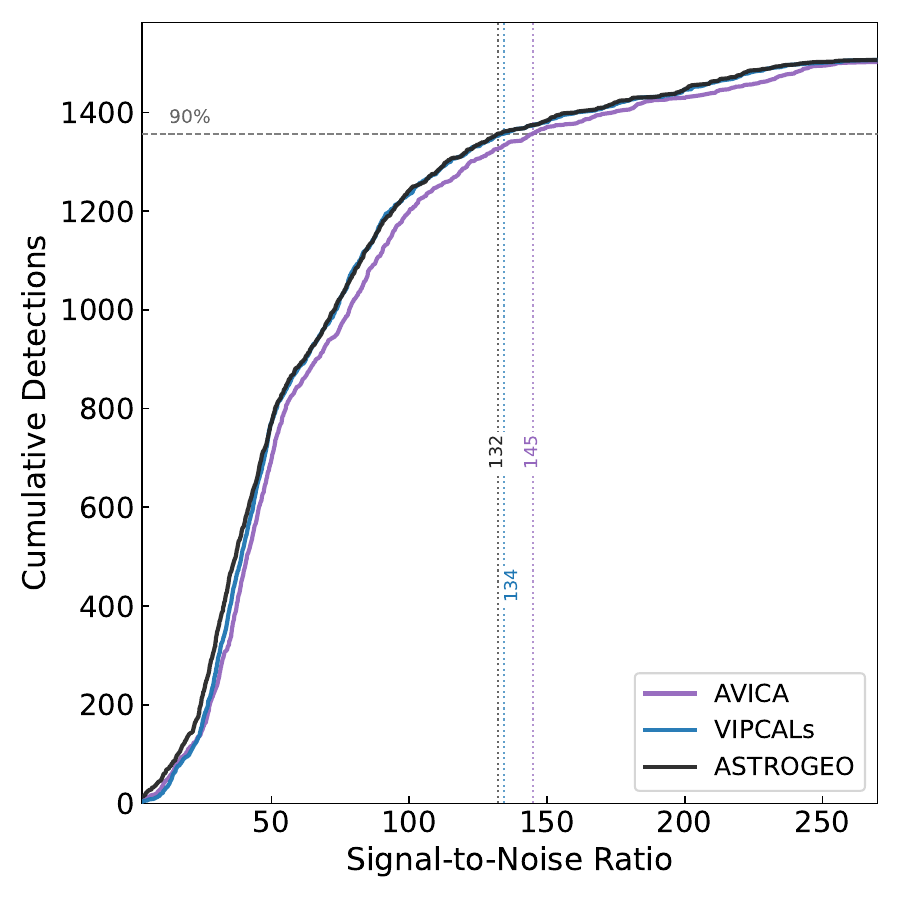}
    \caption{Cumulative S/N distributions for the pooled detections of the 11 \texttt{Astrogeo} datasets with identical detection counts across all three pathways (\texttt{0639+352} excluded), for \texttt{AVICA} (purple), \texttt{VIPCALs} (blue), and the manual \texttt{Astrogeo} reference (black), for a total of 1507 pooled detections. The dashed horizontal line marks the $90\%$ cumulative detection level, at which \texttt{AVICA} reaches $\mathrm{S/N}\,145$ compared to $134$ (\texttt{VIPCALs}) and $132$ (\texttt{Astrogeo}).}
\label{fig:cdf_combined}
\end{figure}
 
The per-source comparisons can be summarized by pooling the detections into a single combined distribution for each pathway, shown as cumulative distributions in Fig.~\ref{fig:cdf_combined}. To ensure a strictly matched comparison, only the 11 datasets with identical detection counts across all three pathways are pooled. \texttt{0639+352} is excluded due to the difference in the total number of detections. The three pooled curves track each other closely, though \texttt{AVICA} shows a slightly larger offset from the manual reference than \texttt{VIPCALs} does. Across the $\mathrm{S/N}\sim50$--$150$ the \texttt{AVICA} curve lies marginally below the other two, indicating a detection population shifted slightly toward higher S/N. At the $90\%$ cumulative detection threshold, \texttt{AVICA} reaches $\mathrm{S/N}\,145$, compared to $134$ for \texttt{VIPCALs} and $132$ for \texttt{Astrogeo}, quantifying this shift. \texttt{VIPCALs} and \texttt{Astrogeo} track each other closely at every threshold, consistent with their close agreement in Table~\ref{table:KStest}.

\subsection{Computational performance}\label{sec:performance}

The pipeline runs for the tested sample were performed on a system equipped with a 32-core Intel Xeon Silver 4314 CPU, 125\,GB of RAM, and 73\,TB of disk space organized in RAID\,6 with Python 3.10.
 
The raw input data for the 1000-source sample, in FITS-IDI format, amounted to a cumulative $19$\,TB. Loading and processing each source produces a set of output products retained by the pipeline: the band-separated, calibrated, and averaged visibility data in MS format, calibration tables, logs, a calibrated UVFITS file and diagnostic plots from \texttt{jplotter}.

The cumulative size of the input MS data across all sources is $10.6$\,TB, and the total size of all other retained output products amounts to $1.38$\,TB.

\begin{figure}[t]
    \centering
    \includegraphics[width=\linewidth]{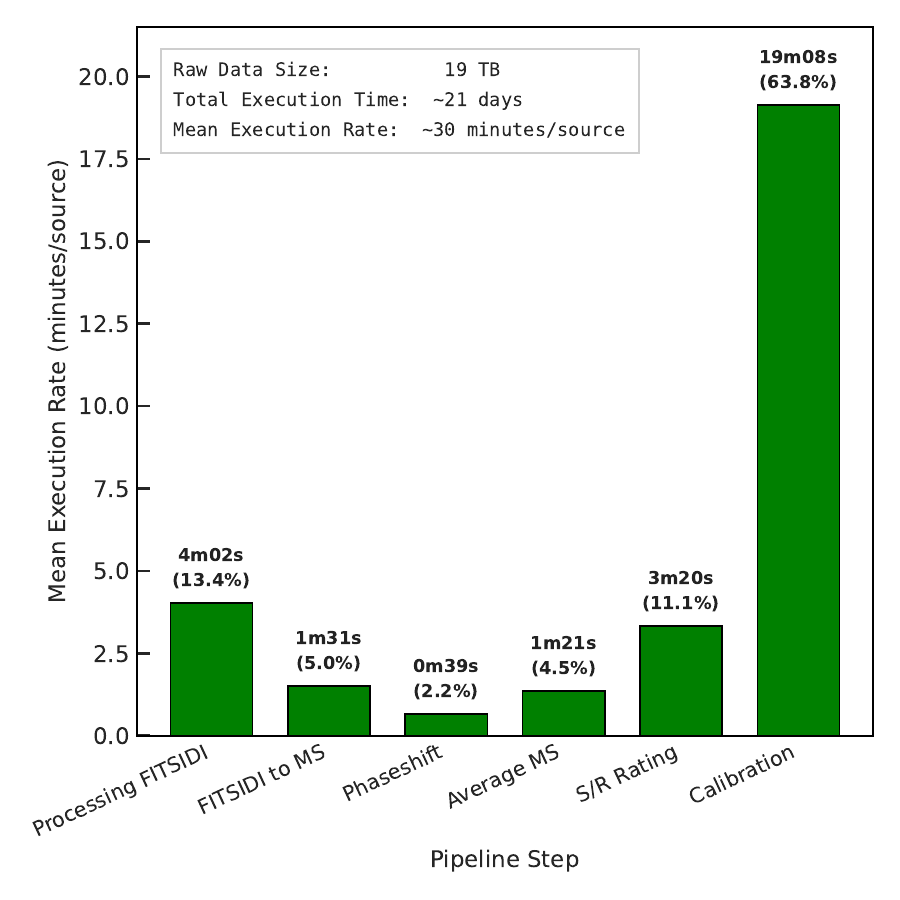}
    \caption{Mean execution rate for each \texttt{AVICA} pipeline step for the 1000-source test sample, expressed in minutes per source. The percentage labels indicate the fractional contribution of each step to the total. Data size: 19\,TB; total pipeline execution time: 21 days.}
    \label{fig:pipelinetime}
\end{figure}

The complete pipeline run with MPI parallelization using up to 20 cores for each dataset concluded in ${\sim}21$\,days, corresponding to a mean per-source runtime of ${\sim}30\,$ minutes. As illustrated in Fig.~\ref{fig:pipelinetime}, the calibration step dominates the total execution time, accounting for 63.8\% of the per-source runtime and amounting to ${\sim}19\,$ minutes for each source. Non-calibration steps such as FITS-IDI preprocessing, data loading, phase shifting, averaging, and S/N rating account for the remaining $36.2\%$ of the per-source execution time.

\begin{figure}[t]
    \centering
    \includegraphics[width=\linewidth]{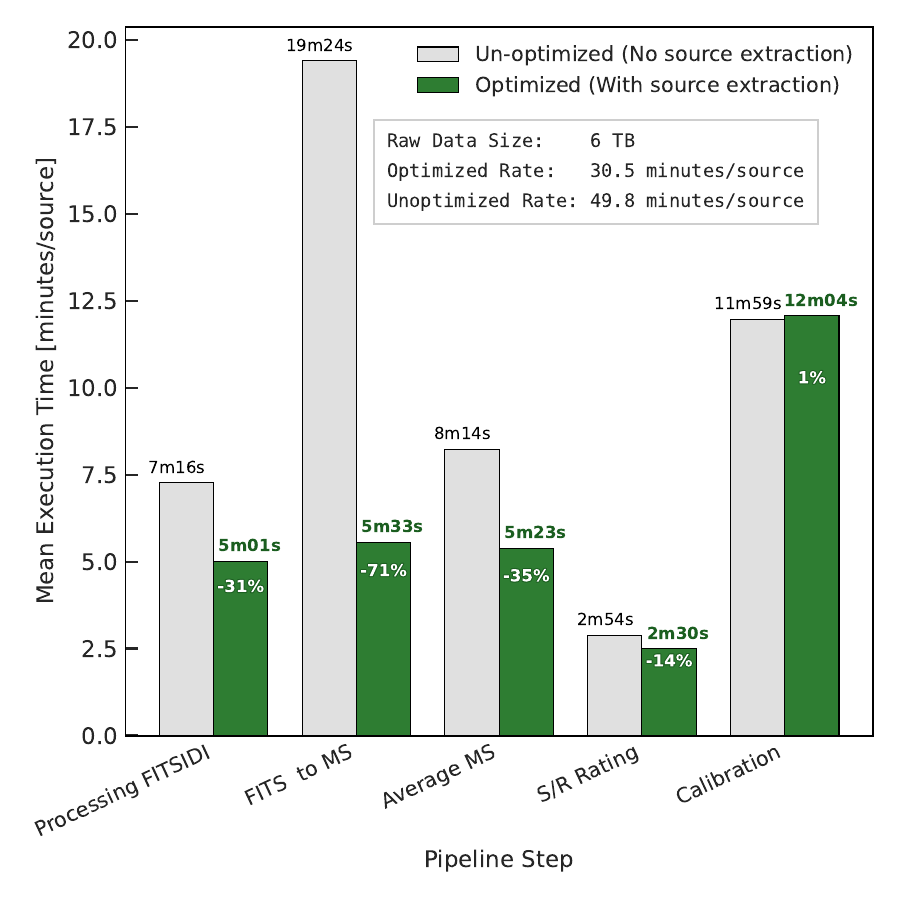}
    \caption{Mean execution time for each pipeline step for \texttt{AVICA} on a subset of 100 sources, comparing two pipeline runs: one with source extraction enabled, in which \textit{fitsidiutil} extracts only the required calibrator and target sources from the FITS-IDI file prior to loading (optimized, green), and one in which the full FITS-IDI file is loaded without prior source extraction (unoptimized, gray).}
    \label{fig:cputime}
\end{figure}

The impact of the source-extraction optimization was assessed on a subset of 100 datasets with a combined size of 6\,TB. Without \textit{fitsidiutil}, the pipeline spent the majority of its time loading the full FITS-IDI files into the MS format, resulting in a mean per-source execution time of ${\sim}50$\,minutes. Extracting only the required sources prior to loading reduced this to ${\sim}30\,$ minutes, an overall improvement of 39\%. As shown in Fig.~\ref{fig:cputime}, the largest gain is in the FITS\,to\,MS conversion step, where the mean time is reduced from 19\,minutes to 5\,minutes per source -- a reduction of 71\%. 

\subsection{Future Developments}\label{sec:future}

 The automated calibrator selection and reference antenna algorithms developed in \texttt{AVICA}, being generalizable to any VLBI array, will be incorporated into a future release of \texttt{rPICARD}, extending its blind-calibration capabilities to the EVN, EHT, and other supported arrays. A calibration feedback tool is planned for future versions of \texttt{AVICA} to estimate the scatter in amplitude and phase for each baseline.

 \texttt{ALFRD} currently supports the creation and simultaneous management of multiple pipeline projects. Planned developments include a fully browser-based graphical interface for pipeline monitoring and control, and a dedicated job scheduler and resource manager to enable scaling of resource-intensive tasks on computing clusters.

\section{Summary}\label{sec:summary}

We have presented \texttt{AVICA}, a modular, fully automated pipeline for the calibration of archival VLBA data within the \texttt{CASA} framework. Developed primarily to process the thousands of diverse VLBA datasets forming the basis of the SMILE project, \texttt{AVICA} utilizes \texttt{ALFRD} as its workflow backend to coordinate pipeline execution, track pipeline status, and visualize progress in real time. The core calibration is performed by \texttt{rPICARD}, which \texttt{AVICA} extends with automated preprocessing at both the FITS-IDI and MS levels, blind selection of calibrators and reference antennas, and comprehensive post-calibration diagnostics.

The pipeline was validated on a sample of 1000 archival VLBA sources, comprising approximately $19\,$TB of data spanning observations from 1995 to 2023. \texttt{AVICA} successfully produced calibrated output for 978 sources ($97.8\%$), with the $22$ failures attributable to corrupted or incomplete archival data. 

Comparison with \texttt{VIPCALs} on the same 1000-source sample demonstrates that when a common S/N threshold of 5.0 is applied, the two pipelines produce consistent fringe solution distributions for the majority of observations, demonstrating that \texttt{AVICA} provides a reliable \texttt{CASA}-based alternative for the fully automated calibration of large-volume, heterogeneous VLBI datasets.

Validation against 12 X-band datasets manually calibrated by experienced observers in the \texttt{Astrogeo} database shows that \texttt{AVICA} and \texttt{VIPCALs} are in close mutual agreement, with effect size $D\leq0.18$ in every case. Both also reproduce the benchmark set by manual calibration to within $D\lesssim0.25$ in all but one case. \texttt{AVICA} shows a slight but consistent tendency toward higher S/N detections relative to the manual benchmark. The effect sizes nonetheless remain small throughout, confirming that automated calibration reliably reproduces the detection statistics of expert manual reduction.

Preprocessing optimizations, specifically the selective extraction of required sources from large FITS-IDI files, reduced the mean per-source execution time from $\sim50\,$ minutes to $\sim30\,$ minutes, an improvement of approximately $39\%$. The fully automated selection of calibrators and reference antennas further eliminates the need for manual parameter input, making the large-volume processing of diverse archival datasets feasible.

By combining a modular architecture with the \texttt{ALFRD} workflow manager, \texttt{AVICA} establishes a standardized, reproducible pipeline capable of processing vast, heterogeneous archival data volumes. The pipeline is publicly available as an open-source Python package.

\begin{acknowledgements}
The authors thank the anonymous referee for their constructive comments, which encouraged a more thorough discussion of calibration-quality assessment. A.K., C.C., D.A. \& F.P. acknowledge support from the European Research Council (ERC) under the Horizon ERC Grants 2021 program under grant agreement No. 101040021. The authors thank Dmitry Blinov and Alejandro Mus for helpful discussions and suggestions. We wish to thank the "Summer School for Astrostatistics in Crete" for providing training on the statistical methods adopted in this work.\\
\end{acknowledgements}

\section*{Data availability}
The \texttt{AVICA} source code is publicly available at \url{https://github.com/avikhagol/avica}.

\bibliographystyle{aa}
\bibliography{mybib.bib}

@ARTICLE{schwabandcotton1983,
       author = {{Schwab}, F.~R. and {Cotton}, W.~D.},
        title = "{Global fringe search techniques for VLBI}",
      journal = {\aj},
         year = 1983,
        month = may,
       volume = {88},
        pages = {688-694},
          doi = {10.1086/113360},
       adsurl = {https://ui.adsabs.harvard.edu/abs/1983AJ.....88..688S}
}

@ARTICLE{VIPCALS2025,
       author = {{{\'A}lvarez-Ortega}, D. and {Casadio}, C. and {P{\"o}tzl}, F.~M. and {Kumar}, A. and {Janssen}, M.},
        title = "{VIPCALs: A fully automated calibration pipeline for very long baseline interferometry data}",
      journal = {\aap},
         year = 2025,
        month = nov,
       volume = {703},
          eid = {A196},
        pages = {A196},
          doi = {10.1051/0004-6361/202556838},
archivePrefix = {arXiv},
       eprint = {2508.13282},
 primaryClass = {astro-ph.IM},
       adsurl = {https://ui.adsabs.harvard.edu/abs/2025A&A...703A.196A}
}

@ARTICLE{Reynolds2002,
       author = {{Reynolds}, C. and {Paragi}, Z. and {Garrett}, M.},
        title = "{Pipeline Processing of VLBI Data}",
      journal = {arXiv e-prints},
         year = 2002,
        month = may,
          eid = {astro-ph/0205118},
        pages = {astro-ph/0205118},
          doi = {10.48550/arXiv.astro-ph/0205118},
archivePrefix = {arXiv},
       eprint = {astro-ph/0205118},
 primaryClass = {astro-ph},
       adsurl = {https://ui.adsabs.harvard.edu/abs/2002astro.ph..5118R}
}

@ARTICLE{2025Petrov&Kovalev,
       author = {{Petrov}, L.~Y. and {Kovalev}, Y.~Y.},
        title = "{The Radio Fundamental Catalog. I. Astrometry}",
      journal = {\apjs},
         year = 2025,
        month = feb,
       volume = {276},
       number = {2},
          eid = {38},
        pages = {38},
          doi = {10.3847/1538-4365/ad8c36},
archivePrefix = {arXiv},
       eprint = {2410.11794},
 primaryClass = {astro-ph.IM},
       adsurl = {https://ui.adsabs.harvard.edu/abs/2025ApJS..276...38P}
}

@ARTICLE{Janssen2019,
       author = {{Janssen}, M. and {Goddi}, C. and {van Bemmel}, I.~M. and {Kettenis}, M. and {Small}, D. and {Liuzzo}, E. and {Rygl}, K. and {Mart{\'\i}-Vidal}, I. and {Blackburn}, L. and {Wielgus}, M. and {Falcke}, H.},
        title = "{rPICARD: A CASA-based calibration pipeline for VLBI data. Calibration and imaging of 7 mm VLBA observations of the AGN jet in M 87}",
      journal = {\aap},
         year = 2019,
        month = jun,
       volume = {626},
          eid = {A75},
        pages = {A75},
          doi = {10.1051/0004-6361/201935181},
archivePrefix = {arXiv},
       eprint = {1905.01905},
 primaryClass = {astro-ph.IM},
       adsurl = {https://ui.adsabs.harvard.edu/abs/2019A&A...626A..75J}
}

@ARTICLE{bemmelCASAonfringe2022,
       author = {{van Bemmel}, Ilse M. and {Kettenis}, Mark and {Small}, Des and {Janssen}, Michael and {Moellenbrock}, George A. and {Petry}, Dirk and {Goddi}, Ciriaco and {Linford}, Justin D. and {Rygl}, Kazi L.~J. and {Liuzzo}, Elisabetta and {Marcote}, Benito and {Bayandina}, Olga S. and {Schweighart}, Neal and {Verkouter}, Marjolein and {Keimpema}, Aard and {Szomoru}, Arpad and {van Langevelde}, Huib Jan},
        title = "{CASA on the Fringe-Development of VLBI Processing Capabilities for CASA}",
      journal = {\pasp},
         year = 2022,
        month = nov,
       volume = {134},
       number = {1041},
          eid = {114502},
        pages = {114502},
          doi = {10.1088/1538-3873/ac81ed},
archivePrefix = {arXiv},
       eprint = {2210.02275},
 primaryClass = {astro-ph.IM},
       adsurl = {https://ui.adsabs.harvard.edu/abs/2022PASP..134k4502V}
}

@INPROCEEDINGS{McMullin2007,
       author = {{McMullin}, J.~P. and {Waters}, B. and {Schiebel}, D. and {Young}, W. and {Golap}, K.},
        title = "{CASA Architecture and Applications}",
    booktitle = {Astronomical Data Analysis Software and Systems XVI},
         year = 2007,
       editor = {{Shaw}, R.~A. and {Hill}, F. and {Bell}, D.~J.},
       series = {Astronomical Society of the Pacific Conference Series},
       volume = {376},
        month = oct,
        pages = {127},
       adsurl = {https://ui.adsabs.harvard.edu/abs/2007ASPC..376..127M}
}

@techreport{Small2022,
    author = {{Small}, D. and {Moellenbrock}, G.},
    title = "{CASA Memo 12: The Fringefit Task in CASA}",
    institution = {National Radio Astronomy Observatory},
    year = 2022
}

@ARTICLE{deller2007,
       author = {{Deller}, A.~T. and {Tingay}, S.~J. and {Bailes}, M. and {West}, C.},
        title = "{DiFX: A Software Correlator for Very Long Baseline Interferometry Using Multiprocessor Computing Environments}",
      journal = {\pasp},
         year = 2007,
        month = mar,
       volume = {119},
       number = {853},
        pages = {318-336},
          doi = {10.1086/513572},
archivePrefix = {arXiv},
       eprint = {astro-ph/0702141},
 primaryClass = {astro-ph},
       adsurl = {https://ui.adsabs.harvard.edu/abs/2007PASP..119..318D}
}

@ARTICLE{Smirnov2011p1,
       author = {{Smirnov}, O.~M.},
        title = "{Revisiting the radio interferometer measurement equation. I. A full-sky Jones formalism}",
      journal = {\aap},
         year = 2011,
        month = mar,
       volume = {527},
          eid = {A106},
        pages = {A106},
          doi = {10.1051/0004-6361/201016082},
archivePrefix = {arXiv},
       eprint = {1101.1764},
 primaryClass = {astro-ph.IM},
       adsurl = {https://ui.adsabs.harvard.edu/abs/2011A&A...527A.106S}
}

@ARTICLE{Hamaker1996,
       author = {{Hamaker}, J.~P. and {Bregman}, J.~D. and {Sault}, R.~J.},
        title = "{Understanding radio polarimetry. I. Mathematical foundations.}",
      journal = {\aaps},
         year = 1996,
        month = may,
       volume = {117},
        pages = {137-147},
       adsurl = {https://ui.adsabs.harvard.edu/abs/1996A&AS..117..137H}
}

@article{martividal2010,
	author = {{Mart{\'i}-Vidal}, I.},
	title = {Optimum estimate of delays and dispersive effects   in low-frequency interferometric observations},
	DOI= "10.1051/0004-6361/200913951",
	url= "https://doi.org/10.1051/0004-6361/200913951",
	journal = {A\&A},
	year = 2010,
	volume = 517,
	pages = "A83",
	month = "",
}

@INPROCEEDINGS{Benson1995,
       author = {{Benson}, J.~M.},
        title = "{The VLBA Correlator}",
    booktitle = {Very Long Baseline Interferometry and the VLBA},
         year = 1995,
       editor = {{Zensus}, J.~A. and {Diamond}, P.~J. and {Napier}, P.~J.},
       series = {Astronomical Society of the Pacific Conference Series},
       volume = {82},
        month = jan,
        pages = {117},
       adsurl = {https://ui.adsabs.harvard.edu/abs/1995ASPC...82..117B}
}

@BOOK{Thompson2017,
       author = {{Thompson}, A. Richard and {Moran}, James M. and {Swenson}, Jr., George W.},
        title = "{Interferometry and Synthesis in Radio Astronomy, 3rd Edition}",
         year = 2017,
         publisher = {Springer International Publishing},
         address = {Cham :},
    edition = {3rd ed. 2017.},
    	url = {http://dx.doi.org/10.1007/978-3-319-44431-4},
          doi = {10.1007/978-3-319-44431-4},
       adsurl = {https://ui.adsabs.harvard.edu/abs/2017isra.book.....T}
}

@ARTICLE{CASA2022,
  author = {{The CASA Team} and others},
  title = {CASA, the Common Astronomy Software Applications for Radio Astronomy},
  journal = {Publications of the Astronomical Society of the Pacific},
  volume = {134},
  pages = {114501},
  year = {2022},
  doi = {10.1088/1538-3873/ac9642}
}

@ARTICLE{Hodgson2016,
       author = {{Hodgson}, Jeffrey A. and {Lee}, Sang-Sung and {Zhao}, Guang-Yao and {Algaba}, Juan-Carlos and {Yun}, Youngjoo and {Jung}, Taehyun and {Byun}, Do-Young},
        title = "{The Automatic Calibration of Korean VLBI Network Data}",
      journal = {Journal of Korean Astronomical Society},
         year = 2016,
        month = aug,
       volume = {49},
       number = {4},
        pages = {137-144},
          doi = {10.5303/JKAS.2016.49.4.137},
archivePrefix = {arXiv},
       eprint = {1607.07969},
 primaryClass = {astro-ph.IM},
       adsurl = {https://ui.adsabs.harvard.edu/abs/2016JKAS...49..137H}
}

@ARTICLE{Blackburn2019,
       author = {{Blackburn}, Lindy and {Chan}, Chi-kwan and {Crew}, Geoffrey B. and {Fish}, Vincent L. and {Issaoun}, Sara and {Johnson}, Michael D. and {Wielgus}, Maciek and {Akiyama}, Kazunori and {Barrett}, John and {Bouman}, Katherine L. and {Cappallo}, Roger and {Chael}, Andrew A. and {Janssen}, Michael and {Lonsdale}, Colin J. and {Doeleman}, Sheperd S.},
        title = "{EHT-HOPS Pipeline for Millimeter VLBI Data Reduction}",
      journal = {\apj},
         year = 2019,
        month = sep,
       volume = {882},
       number = {1},
          eid = {23},
        pages = {23},
          doi = {10.3847/1538-4357/ab328d},
archivePrefix = {arXiv},
       eprint = {1903.08832},
 primaryClass = {astro-ph.IM},
       adsurl = {https://ui.adsabs.harvard.edu/abs/2019ApJ...882...23B}
}

@software{Chael2018,
       author = {{Chael}, Andrew and {Bouman}, Katie and {Johnson}, Michael and {Blackburn}, Lindy and {Shiokawa}, Hotaka},
        title = "{eht-imaging: tools for imaging and simulating VLBI data}",
         year = 2018,
        month = feb,
          eid = {10.5281/zenodo.1173414},
          doi = {10.5281/zenodo.1173414},
      version = {1.0},
    publisher = {Zenodo},
       adsurl = {https://ui.adsabs.harvard.edu/abs/2018zndo...1173414C}
}

@INPROCEEDINGS{Greisen2003,
       author = {{Greisen}, E.~W.},
        title = "{AIPS, the VLA, and the VLBA}",
    booktitle = {Information Handling in Astronomy - Historical Vistas},
         year = 2003,
       editor = {{Heck}, Andr{\'e}},
       series = {Astrophysics and Space Science Library},
       volume = {285},
        month = mar,
        pages = {109},
          doi = {10.1007/0-306-48080-8_7},
       adsurl = {https://ui.adsabs.harvard.edu/abs/2003ASSL..285..109G}
}

@INPROCEEDINGS{Shepherd1997,
       author = {{Shepherd}, M.~C.},
        title = "{Difmap: an Interactive Program for Synthesis Imaging}",
    booktitle = {Astronomical Data Analysis Software and Systems VI},
         year = 1997,
       editor = {{Hunt}, Gareth and {Payne}, Harry},
       series = {Astronomical Society of the Pacific Conference Series},
       volume = {125},
        month = jan,
        pages = {77},
       adsurl = {https://ui.adsabs.harvard.edu/abs/1997ASPC..125...77S}
}

@ARTICLE{Virtanen2020,
       author = {{Virtanen}, Pauli and {Gommers}, Ralf and {Oliphant}, Travis E. and {Haberland}, Matt and {Reddy}, Tyler and {Cournapeau}, David and {Burovski}, Evgeni and {Peterson}, Pearu and {Weckesser}, Warren and {Bright}, Jonathan and {van der Walt}, St{\'e}fan J. and {Brett}, Matthew and {Wilson}, Joshua and {Millman}, K. Jarrod and {Mayorov}, Nikolay and {Nelson}, Andrew R.~J. and {Jones}, Eric and {Kern}, Robert and {Larson}, Eric and {Carey}, C.~J. and {Polat}, {\.I}lhan and {Feng}, Yu and {Moore}, Eric W. and {VanderPlas}, Jake and {Laxalde}, Denis and {Perktold}, Josef and {Cimrman}, Robert and {Henriksen}, Ian and {Quintero}, E.~A. and {Harris}, Charles R. and {Archibald}, Anne M. and {Ribeiro}, Ant{\^o}nio H. and {Pedregosa}, Fabian and {van Mulbregt}, Paul and {SciPy 1.  0 Contributors}},
        title = "{SciPy 1.0: fundamental algorithms for scientific computing in Python}",
      journal = {Nature Medicine},
         year = 2020,
        month = feb,
       volume = {17},
        pages = {261-272},
          doi = {10.1038/s41592-019-0686-2},
archivePrefix = {arXiv},
       eprint = {1907.10121},
 primaryClass = {cs.MS},
       adsurl = {https://ui.adsabs.harvard.edu/abs/2020NaMet..17..261V}
}

@ARTICLE{casadio2021,
       author = {{Casadio}, C. and {Blinov}, D. and {Readhead}, A.~C.~S. and {Browne}, I.~W.~A. and {Wilkinson}, P.~N. and {Hovatta}, T. and {Mandarakas}, N. and {Pavlidou}, V. and {Tassis}, K. and {Vedantham}, H.~K. and {Zensus}, J.~A. and {Diamantopoulos}, V. and {Dolapsaki}, K.~E. and {Gkimisi}, K. and {Kalaitzidakis}, G. and {Mastorakis}, M. and {Nikolaou}, K. and {Ntormousi}, E. and {Pelgrims}, V. and {Psarras}, K.},
        title = "{SMILE: Search for MIlli-LEnses}",
      journal = {\mnras},
         year = 2021,
        month = oct,
       volume = {507},
       number = {1},
        pages = {L6-L10},
          doi = {10.1093/mnrasl/slab082},
archivePrefix = {arXiv},
       eprint = {2107.06896},
 primaryClass = {astro-ph.CO},
       adsurl = {https://ui.adsabs.harvard.edu/abs/2021MNRAS.507L...6C}
}

@ARTICLE{Potzl2025,
       author = {{P{\"o}tzl}, F.~M. and {Casadio}, C. and {Kalaitzidakis}, G. and {{\'A}lvarez-Ortega}, D. and {Kumar}, A. and {Missaglia}, V. and {Blinov}, D. and {Janssen}, M. and {Loudas}, N. and {Pavlidou}, V. and {Readhead}, A.~C.~S. and {Tassis}, K. and {Wilkinson}, P.~N. and {Zensus}, J.~A.},
        title = "{SMILE: Discriminating milli-lens systems in a VLBI pilot project}",
      journal = {\aap},
         year = 2025,
        month = mar,
       volume = {695},
          eid = {A169},
        pages = {A169},
          doi = {10.1051/0004-6361/202452340},
archivePrefix = {arXiv},
       eprint = {2409.15229},
 primaryClass = {astro-ph.GA},
       adsurl = {https://ui.adsabs.harvard.edu/abs/2025A&A...695A.169P}
}

@ARTICLE{Kim2023,
       author = {{Kim}, Dae-Won and {Janssen}, Michael and {Krichbaum}, Thomas P. and {Boccardi}, Bia and {MacDonald}, Nicholas R. and {Ros}, Eduardo and {Lobanov}, Andrei P. and {Zensus}, J. Anton},
        title = "{First GMVA observations with the upgraded NOEMA facility: VLBI imaging of BL Lacertae in a flaring state}",
      journal = {\aap},
         year = 2023,
        month = dec,
       volume = {680},
          eid = {L3},
        pages = {L3},
          doi = {10.1051/0004-6361/202348127},
archivePrefix = {arXiv},
       eprint = {2312.05191},
 primaryClass = {astro-ph.HE},
       adsurl = {https://ui.adsabs.harvard.edu/abs/2023A&A...680L...3K}
}

@ARTICLE{JanssenSoftware2022,
       author = {{Janssen}, Michael and {Radcliffe}, Jack F. and {Wagner}, Jan},
        title = "{Software and Techniques for VLBI Data Processing and Analysis}",
      journal = {Universe},
         year = 2022,
        month = oct,
       volume = {8},
       number = {10},
          eid = {527},
        pages = {527},
          doi = {10.3390/universe8100527},
archivePrefix = {arXiv},
       eprint = {2209.06115},
 primaryClass = {astro-ph.IM},
       adsurl = {https://ui.adsabs.harvard.edu/abs/2022Univ....8..527J}
}

@ARTICLE{Myers2003,
       author = {{Myers}, S.~T. and {Jackson}, N.~J. and {Browne}, I.~W.~A. and {de Bruyn}, A.~G. and {Pearson}, T.~J. and {Readhead}, A.~C.~S. and {Wilkinson}, P.~N. and {Biggs}, A.~D. and {Blandford}, R.~D. and {Fassnacht}, C.~D. and {Koopmans}, L.~V.~E. and {Marlow}, D.~R. and {McKean}, J.~P. and {Norbury}, M.~A. and {Phillips}, P.~M. and {Rusin}, D. and {Shepherd}, M.~C. and {Sykes}, C.~M.},
        title = "{The Cosmic Lens All-Sky Survey - I. Source selection and observations}",
      journal = {\mnras},
         year = 2003,
        month = may,
       volume = {341},
       number = {1},
        pages = {1-12},
          doi = {10.1046/j.1365-8711.2003.06256.x},
archivePrefix = {arXiv},
       eprint = {astro-ph/0211073},
 primaryClass = {astro-ph},
       adsurl = {https://ui.adsabs.harvard.edu/abs/2003MNRAS.341....1M}
}

@ARTICLE{Wilkinson2001,
       author = {{Wilkinson}, P.~N. and {Henstock}, D.~R. and {Browne}, I.~W. and {Polatidis}, A.~G. and {Augusto}, P. and {Readhead}, A.~C. and {Pearson}, T.~J. and {Xu}, W. and {Taylor}, G.~B. and {Vermeulen}, R.~C.},
        title = "{Limits on the Cosmological Abundance of Supermassive Compact Objects from a Search for Multiple Imaging in Compact Radio Sources}",
      journal = {\prl},
         year = 2001,
        month = jan,
       volume = {86},
       number = {4},
        pages = {584-587},
          doi = {10.1103/PhysRevLett.86.584},
archivePrefix = {arXiv},
       eprint = {astro-ph/0101328},
 primaryClass = {astro-ph},
       adsurl = {https://ui.adsabs.harvard.edu/abs/2001PhRvL..86..584W}
}

@ARTICLE{DiepenCasaTableMS2015,
       author = {{van Diepen}, G.~N.~J.},
        title = "{Casacore Table Data System and its use in the MeasurementSet}",
      journal = {Astronomy and Computing},
         year = 2015,
        month = sep,
       volume = {12},
        pages = {174-180},
          doi = {10.1016/j.ascom.2015.06.002},
       adsurl = {https://ui.adsabs.harvard.edu/abs/2015A&C....12..174V}
}

@misc{Offringa2010,
       author = {{Offringa}, A.~R.},
        title = "{AOFlagger: RFI Software}",
 howpublished = {Astrophysics Source Code Library, record ascl:1010.017},
         year = 2010,
        month = oct,
          eid = {ascl:1010.017},
archivePrefix = {ascl},
       eprint = {1010.017},
       adsurl = {https://ui.adsabs.harvard.edu/abs/2010ascl.soft10017O},
          key = {Offringa},
         note = {ascl:1010.017}
}

@ARTICLE{Astropy2013,
       author = {{Astropy Collaboration} and {Robitaille}, Thomas P. and {Tollerud}, Erik J. and {Greenfield}, Perry and {Droettboom}, Michael and {Bray}, Erik and {Aldcroft}, Tom and {Davis}, Matt and {Ginsburg}, Adam and {Price-Whelan}, Adrian M. and {Kerzendorf}, Wolfgang E. and {Conley}, Alexander and {Crighton}, Neil and {Barbary}, Kyle and {Muna}, Demitri and {Ferguson}, Henry and {Grollier}, Fr{\'e}d{\'e}ric and {Parikh}, Madhura M. and {Nair}, Prasanth H. and {Unther}, Hans M. and {Deil}, Christoph and {Woillez}, Julien and {Conseil}, Simon and {Kramer}, Roban and {Turner}, James E.~H. and {Singer}, Leo and {Fox}, Ryan and {Weaver}, Benjamin A. and {Zabalza}, Victor and {Edwards}, Zachary I. and {Azalee Bostroem}, K. and {Burke}, D.~J. and {Casey}, Andrew R. and {Crawford}, Steven M. and {Dencheva}, Nadia and {Ely}, Justin and {Jenness}, Tim and {Labrie}, Kathleen and {Lim}, Pey Lian and {Pierfederici}, Francesco and {Pontzen}, Andrew and {Ptak}, Andy and {Refsdal}, Brian and {Servillat}, Mathieu and {Streicher}, Ole},
        title = "{Astropy: A community Python package for astronomy}",
      journal = {\aap},
         year = 2013,
        month = oct,
       volume = {558},
          eid = {A33},
        pages = {A33},
          doi = {10.1051/0004-6361/201322068},
archivePrefix = {arXiv},
       eprint = {1307.6212},
 primaryClass = {astro-ph.IM},
       adsurl = {https://ui.adsabs.harvard.edu/abs/2013A&A...558A..33A}
}

@INPROCEEDINGS{CFITSIO1999,
       author = {{Pence}, William},
        title = "{CFITSIO, v2.0: A New Full-Featured Data Interface}",
    booktitle = {Astronomical Data Analysis Software and Systems VIII},
         year = 1999,
       editor = {{Mehringer}, David M. and {Plante}, Raymond L. and {Roberts}, Douglas A.},
       series = {Astronomical Society of the Pacific Conference Series},
       volume = {172},
        month = jan,
        pages = {487},
       adsurl = {https://ui.adsabs.harvard.edu/abs/1999ASPC..172..487P}
}

@ARTICLE{Savitzky&Golay1964,
       author = {{Savitzky}, A. and {Golay}, M.~J.~E.},
        title = "{Smoothing and differentiation of data by simplified least squares procedures}",
      journal = {Analytical Chemistry},
         year = 1964,
        month = jan,
       volume = {36},
        pages = {1627-1639},
          doi = {10.1021/ac60214a047},
       adsurl = {https://ui.adsabs.harvard.edu/abs/1964AnaCh..36.1627S}
}

\FloatBarrier
\clearpage

\end{document}